\documentclass{aa}
\usepackage{graphicx}
\usepackage{natbib}
\usepackage{amsmath}
\bibpunct{(}{)}{;}{a}{}{,}

\newcommand{\Ab}{\boldsymbol{A}}
\newcommand{\Bb}{\boldsymbol{B}}
\newcommand{\Db}{\boldsymbol{D}}
\newcommand{\fb}{\boldsymbol{f}}
\newcommand{\fbt}{\boldsymbol{\tilde f}}
\newcommand{\fbh}{\boldsymbol{\widehat{f}}}
\newcommand{\Fb}{\boldsymbol{F}}
\newcommand{\gb}{\boldsymbol{g}}
\newcommand{\gbh}{\boldsymbol{\widehat{g}}}
\newcommand{\gbt}{\boldsymbol{\tilde g}}
\newcommand{\Gb}{\boldsymbol{G}}
\newcommand{\Hb}{\boldsymbol{H}}
\newcommand{\Hcb}{\boldsymbol{{\cal H}}}
\newcommand{\Lb}{\boldsymbol{L}}

\newcommand{\Vb}{\boldsymbol{V}}
\newcommand{\Tb}{\boldsymbol{T}}
\newcommand{\Smat}{\boldsymbol{S}}

\newcommand{\Ub}{\boldsymbol{U}}
\newcommand{\Ib}{\boldsymbol{I}}

\newcommand{\pb}{\boldsymbol{p}}
\newcommand{\Pb}{\boldsymbol{P}}
\newcommand{\eb}{\boldsymbol{e}}

\newcommand{\zb}{\boldsymbol{z}}
\newcommand{\zbt}{\boldsymbol{\tilde z}}

\newcommand{\cmat}{{\cal C}}
\newcommand{\fmatb}{\boldsymbol{{\cal F}}}
\newcommand{\cmatb}{\boldsymbol{{\cal C}}}
\newcommand{\Deltab}{\boldsymbol{\Delta}}
\newcommand{\Sigmab}{\boldsymbol{\Sigma}}
\newcommand{\OO}{\boldsymbol{0}}
\newcommand{\rddots}{\tiny
\begin{array}{ccc} & & \cdot \\ & \cdot & \\ \cdot & & \end{array}
}

\begin{document}
\title{Digital Deblurring of CMB Maps: \\ Performance and
Efficient Implementation}

   \author{R. Vio\inst{1}
          \and
            J.G. Nagy\inst{2}
          \and
          L. Tenorio\inst{3}
          \and
            P. Andreani\inst{4}
          \and
            C. Baccigalupi\inst{5}
          \and
          W. Wamsteker\inst{6}
          }

   \offprints{R. Vio}

   \institute{Chip Computers Consulting s.r.l., Viale Don L.~Sturzo 82,
              S.Liberale di Marcon, 30020 Venice, Italy\\
              ESA-VILSPA, Apartado 50727, 28080 Madrid, Spain\\
              \email{robertovio@tin.it}
         \and
                  Department of Mathematics and Computer Science, Emory University,
                        Atlanta, GA 30322, USA. \\
              \email{nagy@mathcs.emory.edu}
         \and
              Department of Mathematical and Computer Sciences, Colorado School
                  of Mines, Golden CO 80401, USA \\
              \email{ltenorio@Mines.EDU}
         \and
              Osservatorio Astronomico di Padova, vicolo dell'Osservatorio 5,
                  35122 Padua, Italy \\
              \email{andreani@pd.astro.it}
         \and
                   SISSA/ISAS, Via Beirut 4, 34014 Trieste, Italy \\
              \email{bacci@sissa.it}
         \and
             ESA-VILSPA, Apartado 50727, 28080 Madrid, Spain\\
             \email{willem.wamsteker@esa.int}
             }

\date{Received: 30 May 2002; accepted: 14 January 2003}

\abstract{Digital deblurring of images is an important problem that arises in multifrequency observations of the
Cosmic Microwave Background (CMB) where, because of the width of the point spread functions (PSF), maps at
different frequencies suffer a different loss of spatial resolution. Deblurring is useful for various reasons:
first, it helps to restore high frequency components lost through the smoothing effect of the instrument's PSF;
second, emissions at various frequencies observed with different resolutions can be better studied on a comparable
resolution; third, some map-based component separation algorithms require maps with similar level of degradation.
Because of computational efficiency, deblurring is usually done in the frequency domain.
But this approach has some limitations as it requires spatial invariance of the PSF, stationarity of the noise,
and is not flexible in the selection of more appropriate boundary conditions. Deblurring in
real space is more flexible but usually not used because of its high computational cost. In this paper
(the first in a series on the subject) we present new
algorithms that allow the use of real space deblurring techniques even for very large images.
In particular, we consider the use of Tikhonov deblurring of noisy maps with applications to {\it PLANCK}.
We provide details for efficient implementations of the algorithms. Their performance is tested
on Gaussian and non-Gaussian simulated CMB maps, and PSFs with both circular and elliptical symmetry.
Matlab code is made available.
\keywords{Methods: data analysis -- Methods: statistical -- Cosmology: cosmic microwave background}
}
\titlerunning{Digital Deblurring of CMB maps}
\authorrunning{R. Vio, J.G. Nagy, L. Tenorio et al.}
\maketitle

\section{Introduction}

During the last decade, observations of the Cosmic Microwave Background (CMB)
anisotropies have progressed significantly.
After the first evidence of CMB intensity fluctuations
measured by the COBE satellite \citep[see][ and references therein]{smoot99},
several balloon-borne and ground based experiments have successfully detected
CMB anisotropies at degree and sub-degree angular scales
\citep{deb02,hal02,lee01,pad01}.
The {\it MAP} satellite \footnote{http://map.gsfc.nasa.gov} currently in
operation will soon provide full sky maps of CMB
anisotropies at about $20$ arcmin. resolution and a sensitivity
of the order of 10 $\mu$K, on a frequency range extending from $22$ to $90~{\rm GHz}$.
This extraordinary experimental enterprise will be followed by the {\it PLANCK}
satellite, scheduled for launch in 2007
\footnote{http://astro.estec.esa.nl/SA-general/Projects/Planck},
that will provide full sky maps of total intensity and polarization
of CMB anisotropy at a few arcmin. resolution, and a sensitivity
of a few $\mu$K, on nine frequency channels ranging from $30$ to $857~{\rm GHz}$.

These new sets of observations will pose new and challenging questions in
data analysis. Methods will have to be developed to process
the large amount of incoming data, and to extract and separate cosmological
information from foreground emissions from extra-Galactic sources as well as from our own Galaxy.

An important question is the creation of sky maps
based on small angular scale time ordered data from all-sky CMB observations of {\it MAP}
and {\it PLANCK}; efficient map-making algorithms based on
a maximum likelihood approach  have been proposed \citep{nat01,bor01,stom02}. Regarding
the separation of emissions coming from different astrophysical processes,
multifrequency observations can be exploited. For example, available prior information
about the signals can be used in a regularised inversion via
Wiener filtering and maximum entropy methods, either on small sky
patches \citep{hob98} or on the whole sky \citep{bou99, sto02}.
It has been shown that under certain independence assumption on the signals,
the map-operating algorithms based on Independent Component Analysis (ICA)
techniques can be applied on sky patches
\citep{bac00} as well as on the whole sky \citep{mai02}.

In this paper we study a different problem. We consider the effects
of the beam of the instrument used to gather the observations.
We present efficient numerical methods to estimate
the emission pattern lost through the degradation of the instrument's point spread function (PSF)
and noise contamination.
This ``deblurring" process may prove very useful in
CMB data analyses: first, it helps recover
high frequencies smoothed out by the instrument's PSF. Second,
a better understanding of sky emissions, from foregrounds in particular, is achieved if
multifrequency sky maps are compared on a common resolution.
Third, some map-based component separation algorithms, such as ICA
\citep{bac00, mai02} require input maps with similar level
of degradation. Furthermore, although the aim of satellite missions such as {\it Planck} and {\it MAP} is to obtain
full sky maps of the CMB, the strength of the CMB over other backgrounds or contaminating
sources will vary over the sky. Therefore, even if some characteristics
of CMB are estimated on full sky maps, it will be convenient to
check these results on smaller sky patches where CMB dominates the other components (e.g., at high
Galactic latitude and at high observing frequency) and data are free from instrumental
and/or observational problems.

Most deblurring of CMB data has been done in frequency space,
\citep[see][]{hob98, sto02}. This approach is computationally efficient but
has some important limitations; it requires the stationarity of
the contaminating noise
and the spatial invariance of the PSF. Furthermore, it implicitly assumes
periodic boundary conditions which, as we show below, is not necessarily the best choice.
Tikhonov regularization provides a more flexible real domain deblurring technique that can
be used when any of these assumptions are not met.

Until recently, Tikhonov deblurring
was avoided because of its computational cost. For example,
for an $N \times N$ pixel image in the frequency domain
one works with $N \times N$ matrices, while in the spatial domain the matrices are of dimension
$N^2 \times N^2$. However, new efficient algorithms that overcome this
problem have made Tikhonov deblurring a competitive alternative to frequency domain techniques.
In this first paper, we present some of these algorithms and show
their good performance not only in regards to their computational cost but also
in regards to their numerical stability, which is an important characteristics
in any ill-posed inversion problem. Furthermore, we will
present some preliminary results concerning the deblurring for the case of spatially varying PSFs.
A detailed treatment of this problem will be provided in a future paper.

The rest of the paper is organized as follows.
In Sects.~\ref{sec:formal} and \ref{sec:numdeb} we provide a formal definition of the map-based deblurring
process, focusing on boundary conditions and numerical issues. In Sect.~\ref{sec:tik}
we discuss a regularization approach that is efficient and flexible
for deblurring noisy maps; results of numerical experiments are provided in Sect. \ref{sec:numerical}.
Our first application to a realistic map is
presented in Sect. \ref{sec:practical}, where we consider a region in the sky
\citep[already addressed by][]{bac00} in which the CMB emission
largely dominates over foregrounds. We will assume observational
conditions, frequencies, angular resolution and noise level
corresponding to the Low Frequency Instrument (LFI) of the {\it PLANCK}
satellite. Further simulations are presented in Sects.~\ref{sec:nongauss} and \ref{sec:variant}
where we consider non-Gaussian random fields and spatially varying PSFs.
In Sect.~\ref{sec:conclusions} we close with final comments and conclusions.

\section{Formalization of the problem} \label{sec:formal}

When a two-dimensional object $f(\xi,\eta)$ is observed through an
optical (linear) system, it is seen as an image $g(x,y)$
\begin{equation} \label{eq:model1}
g(x,y) = \int\limits_{-\infty}^{+\infty} \int\limits_{-\infty}^{+\infty}
h(x,y,\xi,\eta) f(\xi,\eta)~d\xi~d\eta,
\end{equation}
where the function $h(x,y,\xi,\eta)$, called a {\it point-spread function}
(PSF), represents the blurring action of the optical instrument. Model
(\ref{eq:model1}) allows the PSF to vary with position in
both $(x,y)$ and $(\xi,\eta)$ variables ({\it space-variant} PSF). Often, it
is possible to
simplify this model by assuming that the PSF is independent of position ({\it
space-invariant} PSF), and thus
\begin{equation} \label{eq:model2}
g(x,y) = \int\limits_{-\infty}^{+\infty} \int\limits_{-\infty}^{+\infty} h(x-
\xi,y-\eta) f(\xi,\eta)~d\xi~d\eta.
\end{equation}
Another useful simplification occurs when PSF is
separable, which means that it can be written in the form
\begin{equation}
h(x,y,\xi,\eta)=h_1(x,\xi)~h_2(y,\eta),
\end{equation}
for the space-variant PSF, and
\begin{equation}
h(x-\xi,y-\eta)=h_1(x-\xi)~h_2(y-\eta),
\end{equation}
for the space-invariant PSF. In this case models (\ref{eq:model1}),
(\ref{eq:model2}) can be simplified to
\begin{equation} \label{eq:model3}
g(x,y) = \int\limits_{-\infty}^{+\infty} h_1(x,\xi)  \left[ \int\limits_{-
\infty}^{+\infty} h_2(y,\eta)  f(\xi,\eta)~d\eta \right] d\xi,
\end{equation}
and
\begin{equation} \label{eq:model4}
g(x,y) = \int\limits_{-\infty}^{+\infty} h_1(x-\xi)  \left[ \int\limits_{-
\infty}^{+\infty} h_2(y-\eta)  f(\xi,\eta)~d\eta\right] d\xi,
\end{equation}
respectively. We see that separability of the PSF implies independent blurring
along the horizontal and vertical directions.

Models (\ref{eq:model1}), (\ref{eq:model2}),
(\ref{eq:model3}), and (\ref{eq:model4}) are only theoretical. In practical
applications,
we only have discrete noisy observations of the image $g(x,y)$, which
we model as a discrete linear system
\begin{equation} \label{eq:modeld}
\gb = \Hb \fb + \zb.
\end{equation}
Here, $\gb = {\rm vec} (\Gb)$ and $\fb = {\rm vec} (\Fb)$ are one-dimensional,
column arrays containing, respectively, the observed image $\Gb$ and the true images
$\Fb$ in {\it stacked} order \footnote{We recall that
for a $N \times M$ matrix $\Pb$,
${\rm vec}(\Pb) = ( \pb_1^T ~\pb_2^T \ldots ~\pb_M^T )^T$  with $\pb_i$
the i-th column of matrix $\Pb$.},
$\zb$ is an array containing the noise contribution (assumed to be of additive
type), and $\Hb$ is a matrix that represents the discretized blurring operator.

By image deblurring we mean solving for $\fb$ given the linear system
(\ref{eq:modeld}).
This process is not trivial for the following reasons:
\begin{enumerate}
\item Images are available only in a finite bounded region. However, points near the boundary
are affected by data outside the field of view.
\item If the observed image is an $N \times N$ matrix, then $\fb$, $\gb$,
and $\zb$ are $N^2\times 1$ vectors and $\Hb$ is an $N^2 \times N^2$ matrix. This means
that even for small images (e.g., $N=512$), the matrix $\Hb$ can be quite large.
\item The models (\ref{eq:model1}), (\ref{eq:model2}),
(\ref{eq:model3}), and (\ref{eq:model4}) are particular examples of a class of
integral equations which are well known to be ill-posed \citep{win91}. This implies that the
matrix $\Hb$ is severely ill-conditioned and standard techniques that do not take this into
account are likely to fail.
\end{enumerate}

Points (1) and (2) are addressed in Sect. \ref{sec:numdeb}, and point (3) in Sect. \ref{sec:tik}.

\section{Numerical issues in image deblurring} \label{sec:numdeb}
\subsection{Boundary conditions}\label{sec:bc}
To resolve the difficulty of points near the image boundary being affected by
data outside the field of view, we have to impose some
{\it boundary conditions}.
In image processing, one of the following three boundary conditions
is (either explicitly or implicitly) typically made:
\begin{itemize}
\item[$\bullet$] {\it Periodic} boundary conditions imply that the image repeats
in all directions. That is,
we assume the image $X$ has been extracted from a larger image that looks like:
\begin{displaymath}
\begin{array}{ccc}
X & X & X \\
X & X & X \\
X & X & X
\end{array}
\end{displaymath}
\item[$\bullet$] {\it Zero} boundary conditions imply that the scene outside the borders of
the image $X$ are all zero. That is,
we assume the image $X$ has been extracted from a larger image of the form:
\begin{displaymath}
\begin{array}{ccc}
0 & 0 & 0 \\
0 & X & 0 \\
0 & 0 & 0
\end{array}
\end{displaymath}
\item[$\bullet$]
{\it Reflexive} boundary conditions model the scene outside the image
boundaries as a mirror image of the scene inside the boundaries.
That is, we assume the image $X$ has been
extracted from a larger image that
looks like:
\begin{displaymath}
\begin{array}{ccc}
X_{rc} & X_r & X_{rc} \\
X_c & X & X_c \\
X_{rc} & X_r & X_{rc}
\end{array}
\end{displaymath}
where $X_c$ is obtained by ``flipping'' the columns of $X$, $X_r$ is obtained
by ``flipping'' the rows of $X$, and
$X_{rc}$ is obtained by ``flipping'' the rows and columns of $X$.
\end{itemize}
For a spatially invariant PSF, each of these choices imposes
a particular kind of structure on the matrix $\Hb$.  To describe
these structures we need the following notation:
\begin{itemize}
\item[$\bullet$]
An $N \times N$ matrix with entries that are constant on each
diagonal is called a {\em Toeplitz} matrix, and it can be written as:
\begin{equation}
\label{eq:toep}
  \left(
    \begin{array}{ccccc}
        h_0  & h_{-1}  & h_{-2}  & \cdots & h_{1-N} \\
        h_1  & h_0     & h_{-1}  & \cdots & h_{2-N} \\
        h_2  & h_1     & h_0     & \cdots & h_{3-N} \\
      \vdots & \vdots  & \vdots  & \ddots & \vdots  \\
     h_{N-1} & h_{N-2} & h_{N-3} & \cdots & h_0
    \end{array}
  \right)\,.
\end{equation}
\item[$\bullet$]
If an $N \times N$ Toeplitz matrix has the additional property
that each column (and row) is a circular shift of its previous
column (row), then it is called a {\em circulant} matrix and it
has the form:
\begin{equation}
\label{eq:circ}
  \left(
    \begin{array}{ccccc}
        h_0  & h_{N-1} & h_{N-2} & \cdots & h_1     \\
        h_1  & h_0     & h_{N-1} & \cdots & h_2     \\
        h_2  & h_1     & h_0     & \cdots & h_3     \\
      \vdots & \vdots  & \vdots  & \ddots & \vdots  \\
     h_{N-1} & h_{N-2} & h_{N-3} & \cdots & h_0
    \end{array}
  \right)\,.
\end{equation}
\item[$\bullet$]
An $N \times N$ matrix with entries that are constant on each
anti-diagonal is called a {\em Hankel} matrix. It can be written as:
\begin{equation}
\label{eq:hankel}
  \left(
    \begin{array}{ccccc}
        h_0  & h_1     & h_2     & \cdots & h_{N-1} \\
        h_1  & h_2     & h_3     & \cdots & h_N     \\
        h_2  & h_3     & h_4     & \cdots & h_{N+1} \\
      \vdots & \vdots  & \vdots  & \ddots & \vdots  \\
     h_{N-1} & h_N     & h_{N+1} & \cdots & h_{2N-2}
    \end{array}
  \right)\,.
\end{equation}
\end{itemize}
In two-dimensional applications, such as image deblurring, the
matrix $\Hb$ is usually represented in block form.  In this case,
we need to specify the {\em block structure} of the matrix, as well
as the structure of each block.  For example, if a matrix has
the form:
\begin{equation}
\label{eq:BT}
  \Hb_T = \left(
            \begin{array}{ccccc}
              \Hb_0  & \Hb_{-1}  & \Hb_{-2}  & \cdots & \Hb_{1-N} \\
              \Hb_1  & \Hb_0     & \Hb_{-1}  & \cdots & \Hb_{2-N} \\
              \Hb_2  & \Hb_1     & \Hb_0     & \cdots & \Hb_{3-N} \\
              \vdots & \vdots  & \vdots  & \ddots & \vdots  \\
           \Hb_{N-1} & \Hb_{N-2} & \Hb_{N-3} & \cdots & \Hb_0
            \end{array}
          \right)\,,
\end{equation}
where each $\{ \Hb_i \}$ is an $N \times N$ matrix, then we say $\Hb_T$
is a {\em block Toeplitz} matrix.  If, in addition, each $\{ \Hb_i \}$
is itself a Toeplitz matrix, then we say $\Hb_T$ is a
{\em block Toeplitz matrix with Toeplitz blocks}.  Note that we
can combine any number of structures.  In this paper, we need
the following combinations:
\begin{itemize}
\item[ ] BTTB (block Toeplitz with Toeplitz blocks).
\item[ ] BCCB (block circulant with circulant blocks).
\item[ ] BHHB (block Hankel with Hankel blocks).
\item[ ] BTHB (block Toeplitz with Hankel blocks).
\item[ ] BHTB (block Hankel with Toeplitz blocks).
\end{itemize}
With this notation, we can precisely describe the structure of $\Hb$ for each of
the boundary conditions:
\begin{itemize}
\item[$\bullet$]
With periodic boundary conditions $\Hb$ is BCCB.
\item[$\bullet$]
With zero boundary conditions $\Hb$ is BTTB.
\item[$\bullet$]
With reflexive boundary conditions $\Hb$ is a sum of
BTTB, BTHB, BHTB and BHHB matrices.  (Each of these matrices takes into account,
to some extent, contributions from $X$, $X_r$, $X_c$,
and $X_{rc}$, respectively.)
\end{itemize}
Very often the choice of boundary conditions is considered secondary
to the development of an efficient deblurring algorithm. In fact, in many
situations the type of boundary conditions is
implicitly imposed by the algorithm itself. For example, Fourier methods
implicitly assume periodic boundary conditions.
However, an improper selection of boundary conditions may introduce edge effects such
as discontinuities, which appear as a slow decay of Fourier coefficients and Gibbs oscillations
at the jump, which has obvious consequences on nonGaussianity tests
and reliable estimations of the regularization parameter (see below). Consequently, the choice
of the boundary conditions must be considered an integral part of the deblurring operation and not
merely a byproduct of some algorithm.

\subsection{Numerical methods for the solution of large linear systems}

Even for a very large matrix $\Hb$, there are characteristics of this matrix that make finding
a solution of the system (\ref{eq:modeld}) feasible. For example:
\begin{itemize}
\item[$\bullet$] if $\Hb$ is a highly structured matrix, then it is not necessary to store
all its entries. For instance, for a BTTB matrix it
is sufficient to store the the first column
and first row, whereas for a BCCB matrix it is enough to keep the first row.
\item[$\bullet$] significant memory savings can also be achieved if most of the entries of $\Hb$ are zero,
i.e. $\Hb$ is a sparse matrix.
\end{itemize}
Both of these properties can be efficiently exploited to reduce the computational
cost considerably. Further discussion of these issues
can be found in Appendix \ref{sec:Implementation}.

\section{Numerical regularization}\label{sec:tik}

From Eq.~(\ref{eq:model1}) it is clear the the instrument's PSF smooths out
high frequency components of
the signal, and thus high frequency information is lost. An important consequence of this is
the lack of a unique solution of the linear system (\ref{eq:modeld}); any solution
subjected to high frequency perturbations will fit the data $\gb$ equally well.
This makes the problem of recovering the signal
$\fb$ ill-posed. The ill-posedness also affects the stability
of the solutions; a small perturbation of the data may result in a completely different solution.
Mathematically, the system is ill-posed when the singular values of $\Hb$ decay to zero too fast.
The larger the smoothing effect of the PSF the faster the decay of the singular values.
Small singular values lead to solutions which fit the data but have very large energy; a property
that is not easily justified \citep[for other illustrative
examples of ill-posedness see][]{Hansen97,ten01}. To find
a stable meaningful solution we have to use a {\it regularization} method.

Ill-posed problem can be regularized by imposing constraints on the
unknown signal. For example, a bound on the energy
or a smoothness constraint. This class of constraints can be easily
implemented in the classical framework of {\it Tikhonov
regularization} \citep{tik77}. The idea is to find a solution $\fb_{\lambda}$ that minimizes
a weighted combination of data misfit (residual norm) and signal constraint:
\begin{equation} \label{eq:tikhonov}
\fb_{\lambda} = {\rm argmin}\left(\, \Vert\, \Hb \fb - \gb \,\Vert_2^2 + \lambda^2 \Vert\,
\Lb \fb \,\Vert_2^2\, \right),
\end{equation}
where $\lambda$ is a scalar quantity and $\Lb$ is, for example, the identity matrix (for energy bound) or
a discrete derivative operator of some order (for smoothness constraint). The optimal smoothing functional
for deconvolution operators on an infinite domain was determined by \citet{are74} \citep[see also][]{cul79}
who showed that it is completely determined by the decay rate of the Fourier coefficients
of the unknown function. Unfortunately, in real applications
this result cannot be used directly because, in addition to data being available only on a finite
domain, it requires the function one is trying to recover.

In two-dimensional image processing applications, $\Lb$ is often
chosen to be a discretized Laplace operator, which
is defined as the second order partial derivative
\begin{equation}
  \nabla f(x,y) = \frac{\partial^2 f}{\partial x^2} +
                   \frac{\partial^2 f}{\partial y^2}.
\end{equation}
A discrete approximation can be implemented as a convolution
of $\fb$ with the kernel \citep[][ p.353]{Jain89}
\begin{equation}
\left(
\begin{array}{rrr}
0 & -1 & 0 \\ -1 & 4 & -1 \\ 0 & -1 & 0
\end{array}
\right).
\end{equation}
The implementation of this convolution and
the precise form of the matrix $\Lb$  depend on the boundary conditions. In general, we have
\begin{equation} \label{eq:secondd}
\Lb= \left(
\begin{array}{cccccc}
 \Tb_1 & -\Ib   &        &        &        &  \Tb_3 \\
-\Ib   &  \Tb_2 & -\Ib   &        &        &        \\
       & -\Ib   & \ddots & \ddots &        &        \\
       &        & \ddots & \ddots & -\Ib   &        \\
       &        &        & -\Ib   &  \Tb_2 & -\Ib   \\
 \Tb_3 &        &        &        & -\Ib   &  \Tb_1
\end{array}
\right),
\end{equation}
where $\Ib$ is the identity matrix, and the matrices $\Tb_1$, $\Tb_2$ and $\Tb_3$ depend
of the boundary conditions:
\begin{itemize}
\item[$\bullet$]
for periodic boundary conditions $\Tb_3 = -\Ib$,
\begin{equation}
  \Tb_1 = \Tb_2 =
\left(
\begin{array}{rcccr}
 4 &   -1   &        &        &  -1  \\
-1 & \ddots & \ddots &        &      \\
   & \ddots & \ddots & \ddots &      \\
   &        & \ddots & \ddots & -1   \\
-1 &        &        &  -1    &  4
\end{array}
\right),
\end{equation}
\item[$\bullet$]
for zero boundary conditions $\Tb_3 = 0$,
\begin{equation}
\Tb_1 = \Tb_2 =
\left(
\begin{array}{cccc}
 4 &   -1   &        &      \\
-1 & \ddots & \ddots &      \\
     & \ddots & \ddots & -1 \\
     &        & -1     &  4
\end{array}
\right),
\end{equation}
\item[$\bullet$]
for reflexive boundary conditions $\Tb_3 = 0$,
\begin{equation}
\Tb_1 =
\left(
\begin{array}{rrccrr}
 2 &   -1   &        &        &    &   \\
-1 &    3   &  -1    &        &    &     \\
   &   -1   & \ddots & \ddots &    &     \\
   &        & \ddots & \ddots & -1 &  \\
   &        &        &  -1    &  3 & -1 \\
   &        &        &        & -1 &  2
\end{array}
\right),
\end{equation}
\begin{equation}
\Tb_2 =
\left(
\begin{array}{rrccrr}
 3 &   -1   &        &        &    &   \\
-1 &    4   &  -1    &        &    &     \\
   &   -1   & \ddots & \ddots &    &     \\
   &        & \ddots & \ddots & -1 &  \\
   &        &        &  -1    &  4 & -1 \\
   &        &        &        & -1 &  3
\end{array}
\right).
\end{equation}
\end{itemize}
Other operators, such as first order
derivatives, can be used for $\Lb$ \citep{Jahne}.
However, these operators have the disadvantage that they are
typically applied in only one direction (e.g., $x$-derivative
or $y$-derivative), and therefore are not isotropic.
Nonlinear isotropic implementations are possible but they
are computationally more expensive \citep{Jahne}.
In contrast, the Laplacian is an isotropic linear
operator that can be efficiently incorporated into
our algorithms (see Appendix \ref{sec:Implementation}).

In Eq.~(\ref{eq:tikhonov}), the {\it regularization
parameter} $\lambda$ controls the weight given to the signal
constraint relative to the data misfit. A good selection of $\lambda$ is critical.
A large $\lambda$ favours a small solution norm at the cost of a large data misfit,
while a small $\lambda$ leads to data overfit.

We next consider practical implementations to both solve problem
(\ref{eq:tikhonov}) and estimate an ``optimal'' value of $\lambda$.

\begin{table*}[t]
\begin{center}
\begin{tabular}{ccccccccc}
\hline
\hline
\multicolumn{1}{c}{} & \multicolumn{2}{c}{{\rm Wiener}} &  \multicolumn{1}{c}{}
& \multicolumn{3}{c}{{\rm Tikhonov}} \\
\cline{3-3} \cline{5-9} \\
${\rm FWHM\,(arc.min.)}$ & &${\rm rrms\, (\%)}$ & & ${\rm rrms\, (\%)}$ & & $\widehat{\sigma}/\sigma$
& & $\lambda$\\\hline
\hline
$10$ & & $31.29 \pm 0.05$ & & $30.88 \pm 0.06$ & & $1.002 \pm 0.002$ & & $0.73 \pm 0.01$ \\
$14$ & & $33.62 \pm 0.06$ & & $33.07 \pm 0.06$ & & $1.001 \pm 0.002$ & & $0.75 \pm 0.01$ \\
$23$ & & $38.38 \pm 0.08$ & & $37.45 \pm 0.09$ & & $1.001 \pm 0.002$ & & $0.79 \pm 0.02$ \\
$33$ & & $43.29 \pm 0.09$ & & $41.73 \pm 0.10$ & & $1.001 \pm 0.002$ & & $0.79 \pm 0.03$ \\
\hline
\hline
\end{tabular}
\caption{Results of Tikhonov (with reflexive boundary conditions and discrete Laplacian for $\Lb$)
and Wiener deblurring
of a Gaussian random field whose statistical properties are similar to those expected of the CMB sky
observed with four channels of {\it PLANCK}-LFI for beams with circular symmetry.
The field has been contaminated with $100$ different
realizations of a white noise process ($S/N=2$). The dimensions of the map are
$340 \times 340$ pixels, which corresponds to a sky area of about $20^\circ \times 20^\circ$.
The relative {\it root mean square} ({\rm rrms}) is defined as the ratio of
the residual root mean square ({\rm rms}) to the rms of the true signal. The third column shows
the ratio of the noise standard deviation estimate
$\widehat{\sigma}$ defined by (\ref{sigest}) to the true noise standard deviation $\sigma$. For
the Tikhonov method, mean values and dispersions of the GCV estimates of $\lambda$ are also shown.
\label{tbl:sim1}}
\end{center}
\end{table*}
\begin{table*}[t]
\begin{center}
\begin{tabular}{ccccccccc}
\hline
\hline
\multicolumn{1}{c}{} & \multicolumn{2}{c}{{\rm Wiener}} &  \multicolumn{1}{c}{}
& \multicolumn{3}{c}{{\rm Tikhonov}} \\
\cline{3-3} \cline{5-9} \\
${\rm FWHM\,(arc.min.)}$ & &${\rm rrms\, (\%)}$ & & ${\rm rrms\, (\%)}$ & & $\widehat{\sigma}/\sigma$
& & $\lambda$\\
\hline
\hline
$10$ & & $30.97 \pm 0.72$ & & $30.66 \pm 0.66$ & & $0.96 \pm 0.02$ & & $0.74 \pm 0.04$ \\
$14$ & & $33.24 \pm 0.82$ & & $32.80 \pm 0.75$ & & $0.96 \pm 0.02$ & & $0.77 \pm 0.04$ \\
$23$ & & $37.86 \pm 1.00$ & & $37.12 \pm 0.92$ & & $0.95 \pm 0.03$ & & $0.80 \pm 0.05$ \\
$33$ & & $42.54 \pm 1.18$ & & $41.30 \pm 1.07$ & & $0.93 \pm 0.03$ & & $0.81 \pm 0.05$ \\
\hline
\hline
\end{tabular}
\caption{As in Table~\ref{tbl:sim1} with the only difference that a new Gaussian
random field is generated for each simulation.
\label{tbl:sim2}}
\end{center}
\end{table*}
\begin{table*}[t]
\begin{center}
\begin{tabular}{ccccccccc}
\hline
\hline
\multicolumn{1}{c}{} & \multicolumn{2}{c}{{\rm Wiener}} &  \multicolumn{1}{c}{}
& \multicolumn{3}{c}{{\rm Tikhonov}} \\
\cline{3-3} \cline{5-9} \\
${\rm FWHM\,(arc.min.)}$ & &${\rm rrms\, (\%)}$ & & ${\rm rrms\, (\%)}$ & & $\widehat{\sigma}/\sigma$
& & $\lambda$\\
\hline
\hline
$10$ & & $30.25 \pm 0.72$ & & $29.96 \pm 0.68$ & & $0.91 \pm 0.03$ & & $0.73 \pm 0.03$ \\
$14$ & & $32.29 \pm 0.81$ & & $31.90 \pm 0.75$ & & $0.91 \pm 0.03$ & & $0.76 \pm 0.04$ \\
$23$ & & $36.49 \pm 0.98$ & & $35.85 \pm 0.91$ & & $0.90 \pm 0.04$ & & $0.79 \pm 0.04$ \\
$33$ & & $40.76 \pm 1.15$ & & $39.71 \pm 1.04$ & & $0.89 \pm 0.04$ & & $0.81 \pm 0.05$ \\
\hline
\hline
\end{tabular}
\caption{As in Table~\ref{tbl:sim2} but with a PSF of elliptical symmetry. The
first column provides the ${\rm FWHM}$ along the major axis that is $1.3$
times the ${\rm FWHM}$ along the minor axis. Axes of the PSF are parallel to the edges of the maps.
\label{tbl:sim2a}}
\end{center}
\end{table*}
\begin{table*}[t]
\begin{center}
\begin{tabular}{ccccccccccc}
\hline
\hline
& & & & & & $ {\rm rms}_W/ {\rm rms}_T$ & & & & \\
\cline{3-11} \\
${\rm FWHM\,(arc.min.)}$ & & $\beta=1.0$ & & $\beta=0.5$ & & $\beta=0.25$
& & $\beta=0.15$ & & $\beta=0.10$ \\
\hline
\hline
$10$ & & $1.01$ & & $1.03$ & & $1.10$ & & $1.14$ & & $1.16$ \\
$14$ & & $1.02$ & & $1.03$ & & $1.10$ & & $1.14$ & & $1.16$ \\
$23$ & & $1.02$ & & $1.05$ & & $1.11$ & & $1.15$ & & $1.17$ \\
$33$ & & $1.04$ & & $1.07$ & & $1.16$ & & $1.21$ & & $1.23$ \\
\hline
\hline
\end{tabular}
\caption{Ratio ${\rm rms}_W / {\rm rms}_T$ for different values of the
parameter $\beta$ that increases the correlation length (see text).
\label{tbl:wfvstik}}
\end{center}
\end{table*}

\subsection{Numerical methods for Tikhonov map deblurring} \label{sec:tikmap}

As previously mentioned, the image deblurring problem
is severely ill-conditioned and regularization is needed
in order to compute solutions that are not completely corrupted
by noise.  One approach is Tikhonov regularization,
where the solution $\fb_\lambda$ is given by Eq.~(\ref{eq:tikhonov}).
Stable algorithms for computing this solution
are usually developed by reformulating (\ref{eq:tikhonov})
as the damped least squares problem
\begin{equation}
\label{eq:tikls}
  \min_{\fb} \left\|
                \left( \begin{array}{c}
                          \Hb \\ \lambda \Lb
                       \end{array}
                \right)
                \fb -
                \left( \begin{array}{c}
                          \gb \\ \OO
                       \end{array}
                \right)
              \right\|_{2}\,.
\end{equation}
It is important to note that the Tikhonov method is just one
approach that can be used for regularization but many other
schemes may be used.  For large scale image deblurring
problems, iterative regularization methods, such as
conjugate gradients, Landweber iteration,
or expectation-maximization (sometimes referred to as Richardson-Lucy),
are often recommended.  Regularization is enforced through
iteration truncation; that is, the iteration index acts as the
regularization parameter.  Although the specific details of
various iterative methods may be different, they usually have the common
property that the most computationally expensive operation
at each iteration is matrix vector multiplications with $\Hb$.
For large scale image
deblurring problems, these computations can be done very efficiently
using fast Fourier transforms.

A disadvantage with using iterative methods is that it is very
difficult to determine when to stop the iteration; that is, how
to choose a good regularization parameter.  Many methods have been
proposed, but they are usually most reliable when
used in combination with an interactive visualization of the
restorations computed at each iteration.
Unfortunately, visualization of CMB maps does not provide an accurate
assessment of the accuracy of the computed solution.

In the case of Tikhonov regularization,
choosing a regularization parameter $\lambda$ is
also a non-trivial issue.  However, in comparison with
stopping criteria for iterative methods,
much more work has been done in this area
\citep{EnHaNe00,Hansen97,vog02}.  The generalized cross validation (GCV)
method is probably the most well-known scheme for choosing a value
for $\lambda$.
In this scheme, $\lambda$ is chosen to minimize the ${\rm GCV}$ function
\begin{equation}
\label{eq:gcv1}
{\rm GCV}(\lambda)= \frac{||\Hb\fb_\lambda - \gb||^{2}_{2}/n}
 {[\,\rm{trace}(\Ib-\Hcb(\lambda))/n\,]^{2}},
\end{equation}
where $\Hcb$ is the matrix that defines the estimator of $\Hb\fb$,
i.e., $\Hcb\gb = \Hb\fb_{\lambda}$, and $n$ is the number of pixels in the image.
For Tikhonov regularization
\begin{equation}
\Hcb(\lambda) =\Hb (\Hb^{T}\Hb+\lambda^{2}\Lb^T\Lb)^{-1}\Hb^{T}.
\label{hat:eq}
\end{equation}
The development of the GCV method is based on the
assumption that a good value of the
regularization parameter should predict missing data values
\citep[see Appendix \ref{sec:gcv} and][]{GoHeWh79}.

An additional argument for using Tikhonov regularization
with GCV is that both have been well studied, and various
authors have found this combination of regularization
and parameter choice method to be very robust
\citep{ThBrKaTi91,HaHa93,Hansen97,vog02}.
In Appendix \ref{sec:Implementation} we describe how to efficiently compute
both the minimum of the ${\rm GCV}$ function (thus computing the
regularization parameter) and the solution of the
least squares problem (\ref{eq:tikls}).

\section{Numerical experiments} \label{sec:numerical}

We have used Monte Carlo simulations to check the
reliability and performance of the methodology presented in the previous
sections. The simulations have been conducted under two different scenarios.
We first fix the sky and generate different realizations of the noise
process. Then, to account for the variability of the random field,
we simulate different realizations of the random field and of the
noise process.

To keep the conditions of the experiment under control but still
relate it to the CMB problem considered in the next section, we have used Gaussian
random fields characterized by a correlation function that has been obtained from
the correlation function of the CMB via an approximation with an exponential function (for the details
about the parameters used for CMB see Sect.~\ref{sec:practical}).
A dominating CMB component is a good approximation
to real sky maps at least at medium and high Galactic latitudes where the
effects of diffuse foregrounds from our Galaxy can be
neglected \cite[see][ and references therein]{mai02}. The PSFs used in the experiments
are Gaussians with ${\rm FWHM}$ similar to that expected for the {\it PLANCK}-LFI instrument.
However, since the exact form of the observing beam has not yet been determined with sufficient
accuracy, we have tried two different scenarios: PSFs having circular and elliptical symmetry.
For the latter, the ${\rm FWHM}$
along the major axis is set to $1.3$ times the ${\rm FWHM}$ along the minor axis. The axes
of the beams are parallel to the edges of the maps.

Since the random field is Gaussian and stationary and the noise is assumed
white, classical Wiener filtering is expected to provide the smallest mean
square error among linear filters.
However, this filter requires knowledge of the spectrum of the unknown
signal which is not available in practice. We use Wiener deblurring as a
sort of benchmark to assess the performance of Tikhonov methods.

Simulations were done using reflexive, periodic and zero boundary conditions and with
two different types of penalty matrix $\Lb$, the identity matrix and the discrete
approximation of the second derivative operator \citep[note that realizations of a random field
are smooth under mild regularity conditions on the correlation function, e.g., ][]{adler}.
However, since reflexive boundary conditions and the Laplacian operator
have systematically provided better performance, only results concerning this combination are presented.
In particular, we have found that the choice of the boundary conditions is a
critical factor. In fact, not only have the results been systematically worse
with zero and periodic boundary conditions, but with the latter we
have also encountered stability problems in the estimation of the regularization parameter.
These effects are the result of discontinuities introduced by zero and periodic boundary conditions;
a problem that is much less important for reflexive boundary conditions \citep{NgChTa99}.

The results of the simulations are shown in Tabs.~\ref{tbl:sim1} and \ref{tbl:sim2} for circularly
symmetric PSFs, and in Table~\ref{tbl:sim2a} for PSFs with elliptical symmetry.
Note that GCV estimates of the ``optimal''value of $\lambda$
are stable with respect to both noise and different realizations of the random field.
This is an important indication of the reliability of the methodology. It is well known, however,
that GCV estimates may occasionally give very small values of $\lambda$ resulting in an under-smoothed
solution, this problem can be easily corrected using a procedure defined in Appendix \ref{sec:gcv}.

The same tables also show the mean value of the relative root mean square (rrsm) error, defined
as $\Vert\fb-\fb_\lambda\Vert_2/\Vert\fb\Vert_2$. As expected,
the error increases with
the {\rm FWHM} of the beam. The rrms error indicates that Wiener and Tikhonov
deblurring (with reflexive boundary conditions and and discrete Laplacian)
give comparable results. The advantage of the latter is that it does not require the spectrum of the unknown signal.
That is, the smoothness constraint and the spectrum information provide similar results.
(Note that since the Wiener filter has been implemented using the Fourier transform,
only periodic boundary conditions were used for Wiener deblurring.)

The third column in the tables compares estimates of the noise standard deviation obtained using
({\ref{sigest}), a formula which arises
naturally in the Tikhonov framework, to its
true value $\sigma$. These estimates are necessary to construct confidence intervals or test
hypotheses. An over-estimate of $\sigma$ indicates under-smoothing of the signal estimate
while an underestimate indicates over-smoothing.
We see that Tikhonov provides reasonably good estimates of $\sigma$.

An important difference between Wiener and Tikhonov filters is that the former provides
deblurred estimates assuming that $\fb$ is a realization of a process with
a particular spectrum, while the latter
relies on the particular fixed realization of $\fb$ on which the data are based.
In particular, the Wiener filter minimizes an
error over all possible realizations of the signal while the Tikhonov filter
assumes the signal is fixed. This means that while Wiener filtering
relies on knowledge of the spectrum of the process to approximate the optimal
filtering, Tikhonov uses the data to obtain an estimate of the signal's Fourier
transform. Wiener filtering may thus give misleading results when the
available realization of the signal is in the tail of the process
distribution or when the particular realization of $\fb$ happens to be somewhat unusual for the process,
as it may happen when the assumed
spectrum is incorrect. To illustrate this point, we repeated the simulations conducted for Table~\ref{tbl:sim1}
but chose the initial sky realization from a process with successively larger correlation lengths;
that is we incorrectly specified the spectrum required by Wiener filtering by multiplying
the exponent of the correlation function by $\beta$.
Table~\ref{tbl:wfvstik} shows the ratio ${\rm rms}_W / {\rm rms}_T$ of the rms error of
the Wiener and Tikhonov deblurred estimates for different values of $\beta$. The results
show that Tikhonov deblurring may provide better estimates when the spectrum
is incorrectly specified.

\begin{figure}
        \resizebox{\hsize}{!}{\includegraphics{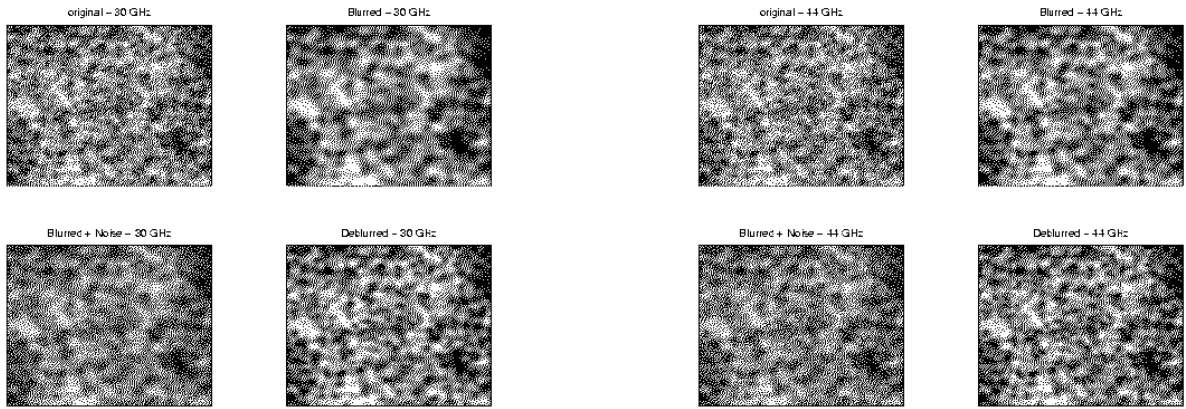}}
        \caption{Grayscale image of the simulated sky maps at $30~{\rm GHz}$ (see text). Each map contains
        $340 \times 340$ square pixels with side of $3.5^{\prime}$ for a total area of about
          $20^\circ \times 20^\circ$.}
        \label{fig:map30}
\end{figure}
\begin{figure}
        \resizebox{\hsize}{!}{\includegraphics{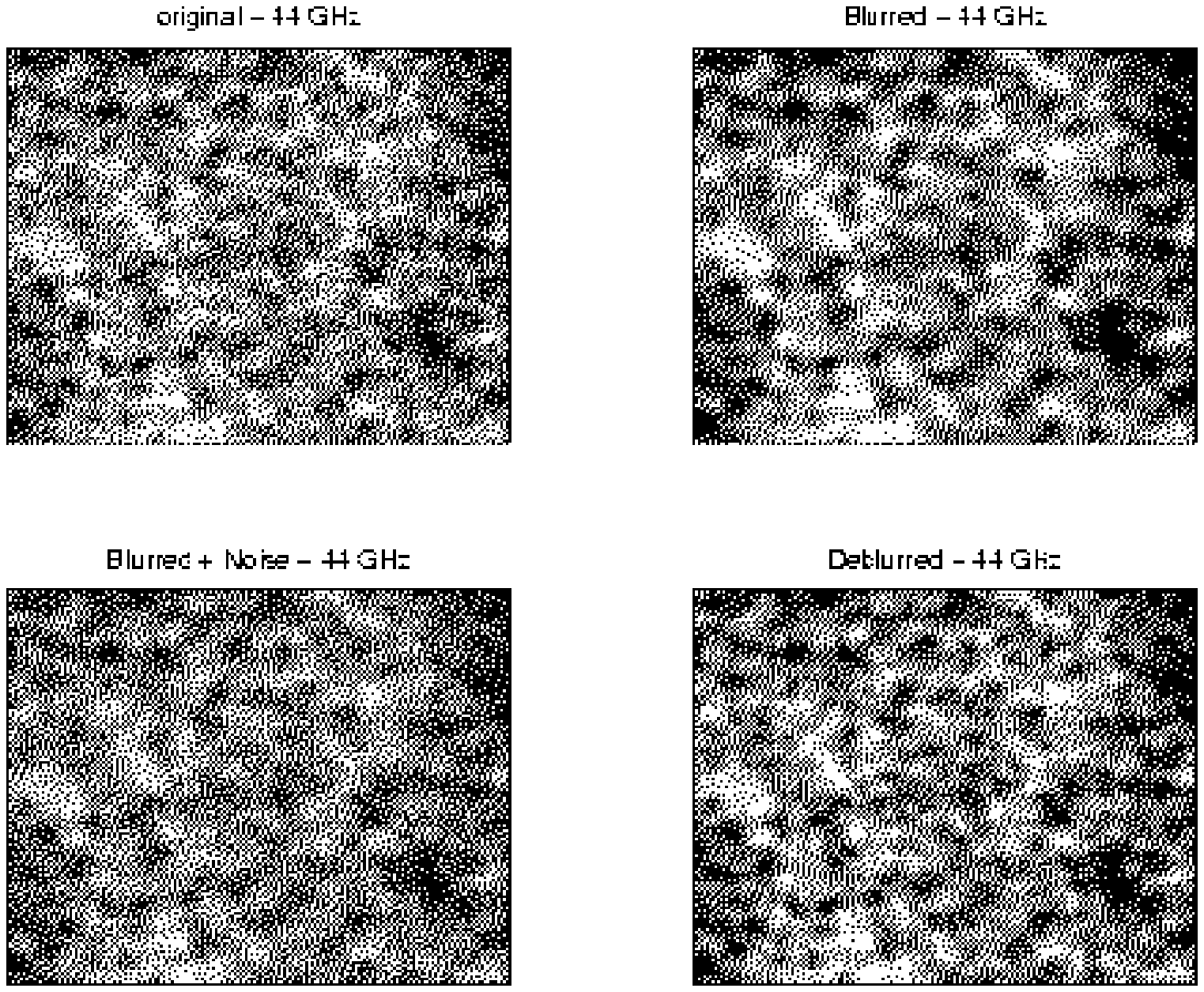}}
        \caption{Same as Fig.~\ref{fig:map30} but at $44~{\rm GHz}$ (see text).}
        \label{fig:map44}
\end{figure}
\begin{figure}
        \resizebox{\hsize}{!}{\includegraphics{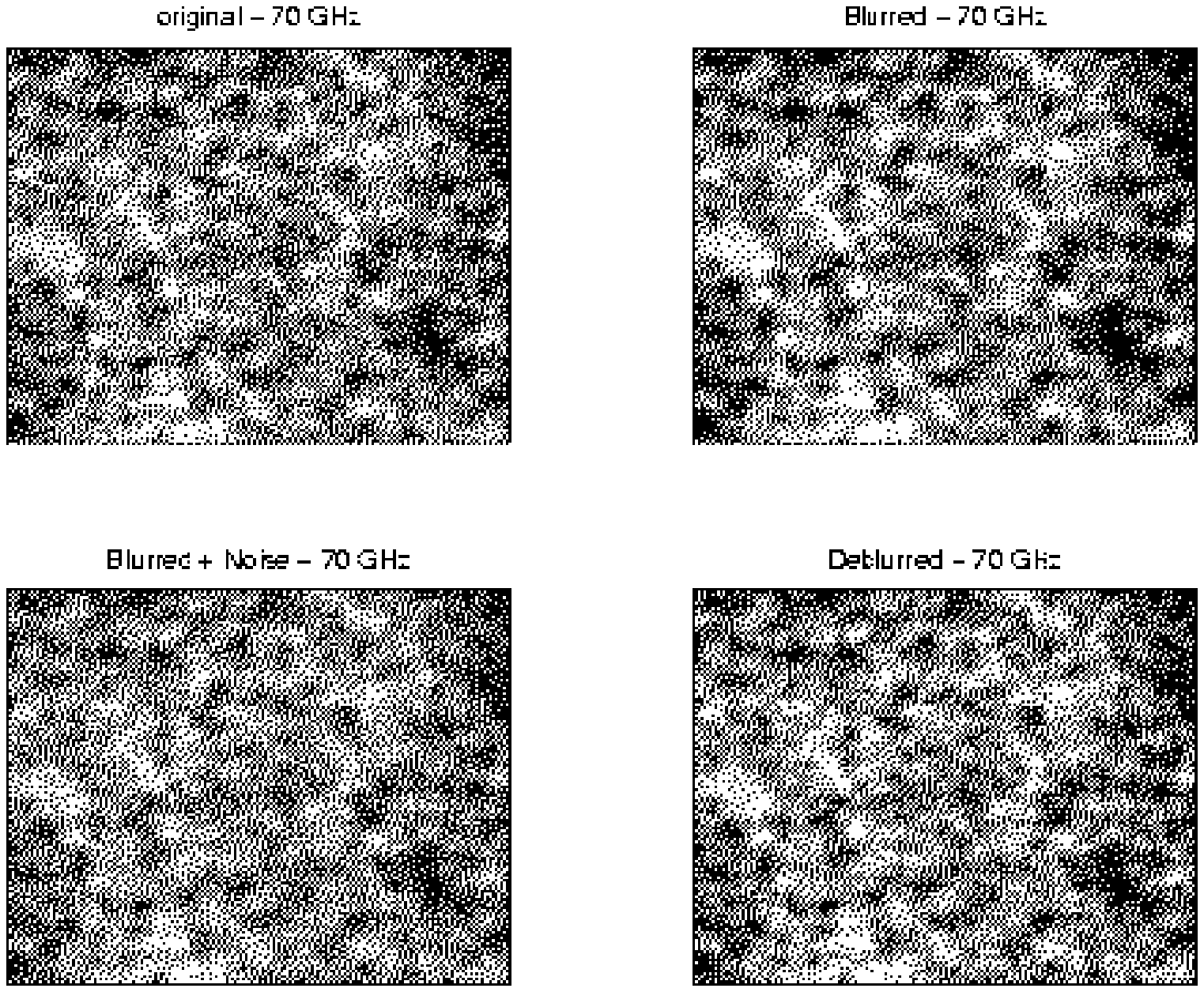}}
        \caption{Same as Fig.~\ref{fig:map30} but at $70~{\rm GHz}$ (see text).}
        \label{fig:map70}
\end{figure}
\begin{figure}
        \resizebox{\hsize}{!}{\includegraphics{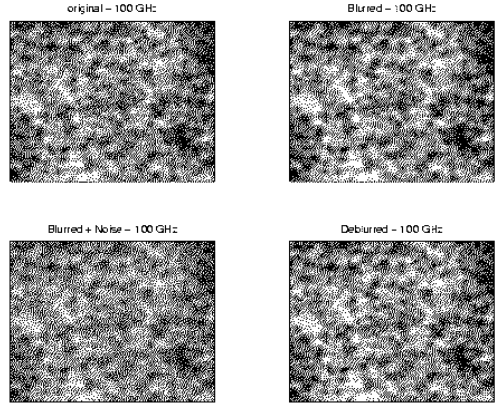}}
        \caption{Same as Fig.~\ref{fig:map30} but at $100~{\rm GHz}$ (see text).}
        \label{fig:map100}
\end{figure}

\section{A CMB application} \label{sec:practical}
We now present tests of the deblurring technique on simulated observations
of the {\it PLANCK}-LFI. The region
analysed is almost identical to the one in \cite{bac00}:
it is a squared patch ($340 \times 340$ pixels) with side
of about $20^{\circ}$, centered
at $l=90^{\circ}$, $b=45^{\circ}$ (Galactic coordinates). The latitude
is high enough that CMB emission dominates over foregrounds, assumed
to be represented by synchrotron \citep{has82} and dust \citep{sch98}
emission. We neglect contributions of point
sources. The CMB model, in agreement with current experimental results
\citep{deb02, hal02, lee01},
corresponds to a flat Friedmann-Robertson-Walker (FRW) metric
with a cosmological constant ($70\%$ of the critical density),
Hubble parameter today $H_{0}=100h$ km/sec/Mpc with $h=0.7$
baryons at $5\%$ and Cold Dark Matter (25\% CDM), with
a scale-invariant Gaussian initial spectrum of adiabatic
density perturbations.

The {\it PLANCK}-LFI instrument works at frequencies
$30$, $44$, $70$, and $100~{\rm GHz}$. We assume nominal
noise and angular resolution
\footnote{http://astro.estec.esa.nl/SA-general/Projects/Planck}.
The maps are blurred through Gaussian PSF's with circular symmetry and
appropriate FWHM's (i.e., $\approx 33^{\prime}$ at $30 ~{\rm GHz}$,
$\approx 23^{\prime}$ at $44 ~{\rm GHz}$,
$\approx 14^{\prime}$ at $70 ~{\rm GHz}$,
$\approx 10^{\prime}$ at $100 ~{\rm GHz}$)
and summed up together with simulated white noise with
{\rm rms} level as expected for the considered channels.
Since we choose to work with a pixel size of about $3.5$ arcminutes, the noise
rms are $.042$, $.049$, $.042$ and $.043$ mK in antenna temperature at $30$, $44$, $70$,
$100~{\rm GHz}$, respectively.

\begin{figure}
        \resizebox{\hsize}{!}{\includegraphics{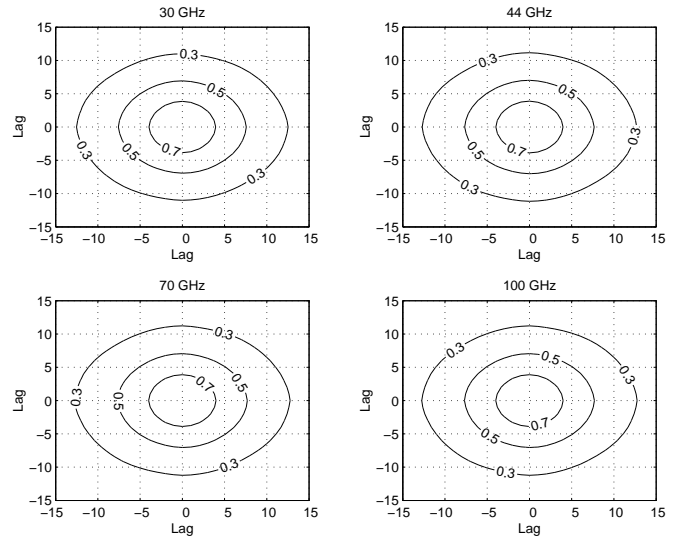}}
        \caption{Contour plot of the two-dimensional auto-correlation function of
the original simulated CMB maps. Lag in pixels.}
        \label{fig:cross1}
\end{figure}
\begin{figure}
        \resizebox{\hsize}{!}{\includegraphics{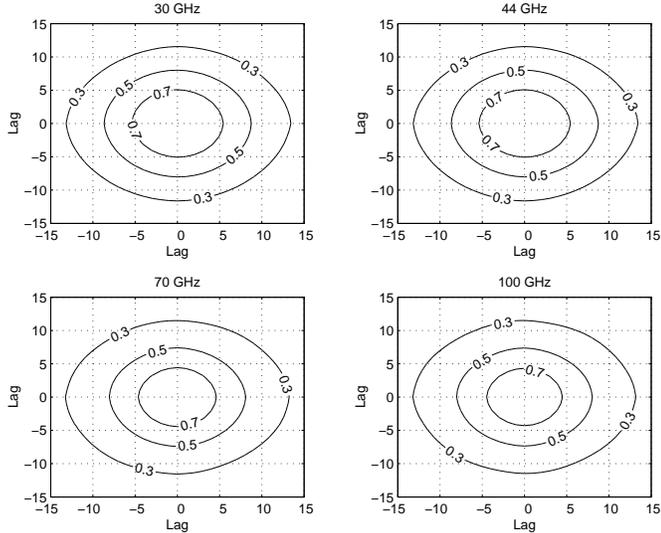}}
        \caption{Contour plot of the two-dimensional cross-correlation function of
          the original with the deblurred CMB maps. Lag in pixels.}
        \label{fig:cross3}
\end{figure}
\begin{figure}
        \resizebox{\hsize}{!}{\includegraphics{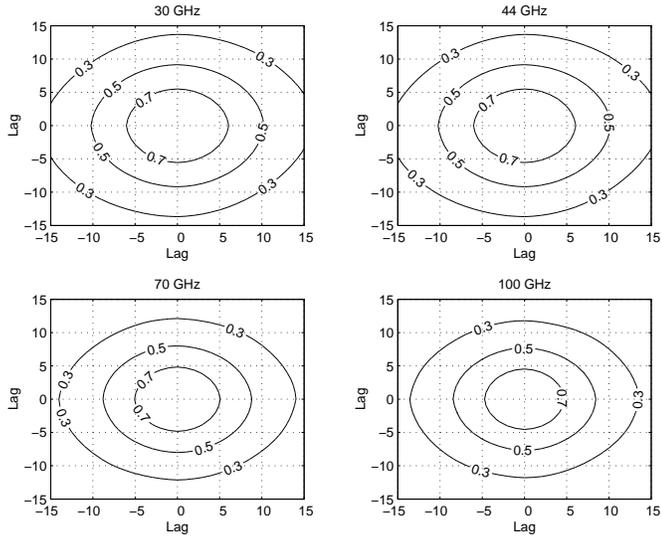}}
        \caption{Contour plot of the two-dimensional cross-correlation function of
          the original with the blurred noise-free CMB maps. Lag in pixels.}
        \label{fig:cross2}
\end{figure}
\begin{figure}
        \resizebox{\hsize}{!}{\includegraphics{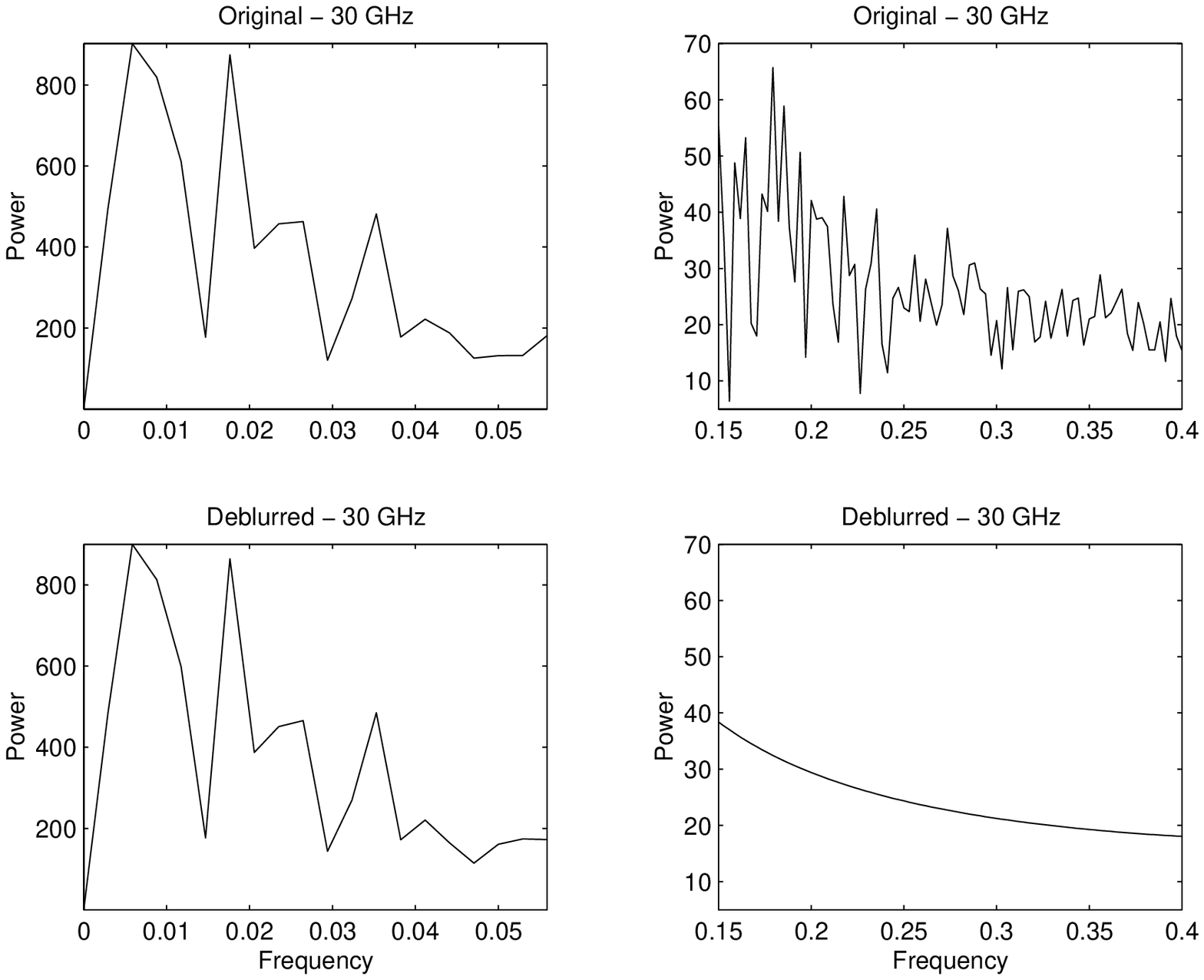}}
        \caption{One-dimensional cross-section of the two-dimensional Power-spectra
          in two different frequency ranges
          of the original $30 ~{\rm GHz}$
        and the deblurred CMB maps (see text). Frequency is in Nyquist units.}
        \label{fig:power}
\end{figure}

The original maps are shown in Figs.~\ref{fig:map30}-\ref{fig:map100}
together with the blurred, blurred plus noise, and deblurred versions (Tikhonov
method with reflexive boundary conditions and discrete Laplacian for $\Lb$).
The deblurred maps look reasonably good
despite the high noise level. But, as expected, there is a clear loss
of high frequencies, especially for the lowest frequency maps. This loss, which is
intrinsic to any deblurring operation, is important given the high noise level.
For example, Figs.~\ref{fig:cross1}-\ref{fig:cross2} show, respectively, the two-dimensional
auto-correlation function of the original maps and their
two-dimensional cross-correlation functions with the blurred (but noise-free) and the deblurred
maps. These figures show that the deblurring operation provides good results for
the lowest contour levels (those mainly determined by the lowest
frequencies) and worse results for the highest contour levels (
those determined mainly by the highest frequencies). The same effect can also be seen in
Fig.~\ref{fig:power}, which shows a one-dimensional cross-section of the power-spectrum
of the original $30 ~{\rm GHz}$ and of the deblurred maps.

The performance of the deblurring procedure can be also checked through the angular
power spectrum, which, as usual, is defined by the expansion coefficients $C_{\ell}$ of the two point
correlation function in Legendre polynomials. Here
$\ell$ is the multipole associated to an angular scale of about $180/\ell$ degrees. Consequently, our analysis
applies to $\ell$ ranging from $\ell\simeq 200$, corresponding to the degree scale
(larger angles are poorly probed due to the finite extension of our patches), up to
multipoles corresponding to the instrumental resolution at each frequency.

Estimates of the $C_{\ell}$ coefficients for the
original, blurred, blurred noisy, and deblurred maps are shown
in Figs.~\ref{fig:ps30}-\ref{fig:ps100}.
Four CMB acoustic
peaks are clearly seen in the original spectrum at $\ell\simeq 200, 500, 800, 1100$.
Since we are
analysing a limited part of the sky, sample variance causes oscillations
in the coefficients, with increasing amplitude as the angular scale
approaches the size of the patch, corresponding to the low multipole tail.

\begin{figure}
        \resizebox{\hsize}{!}{\includegraphics{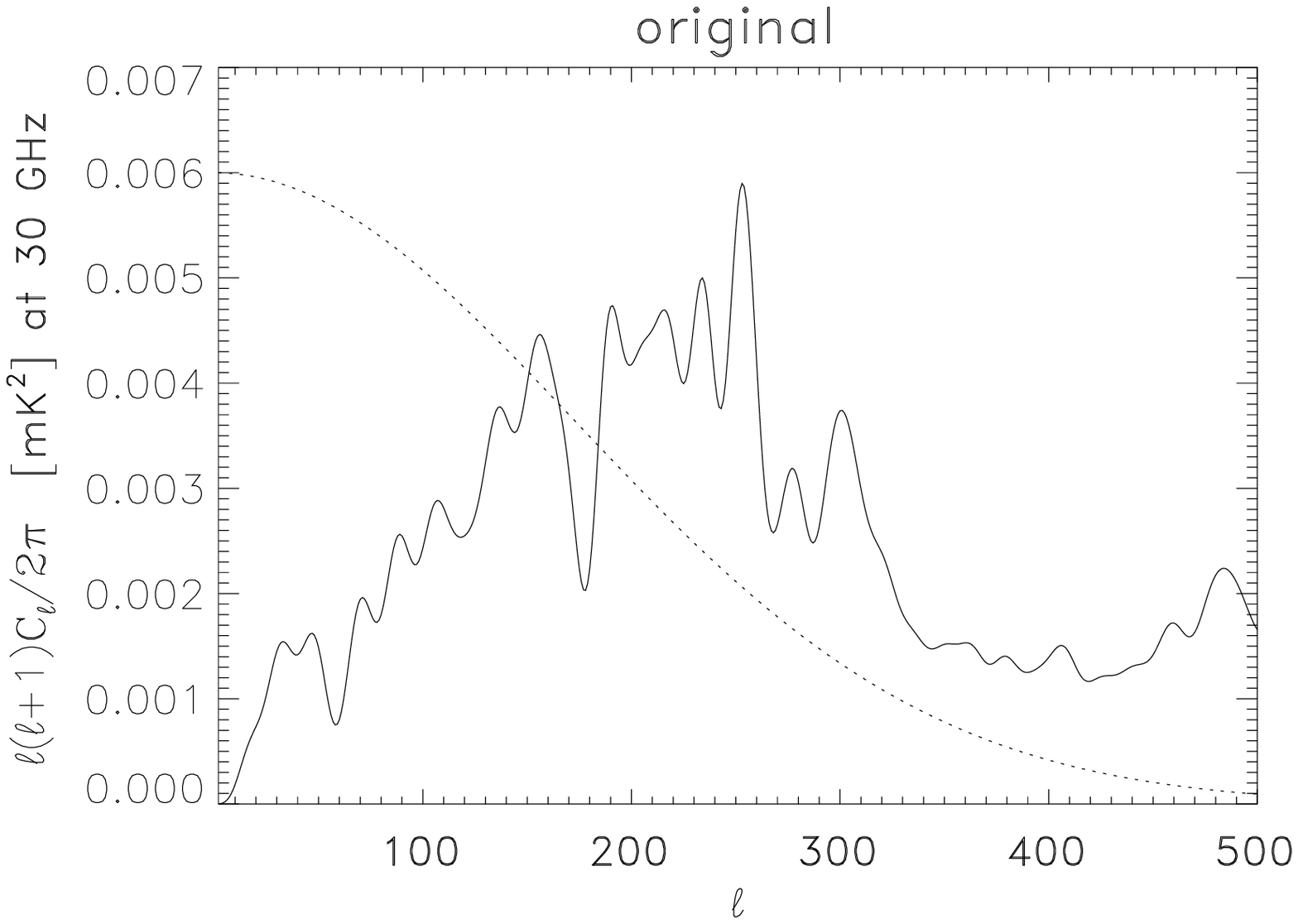}\includegraphics{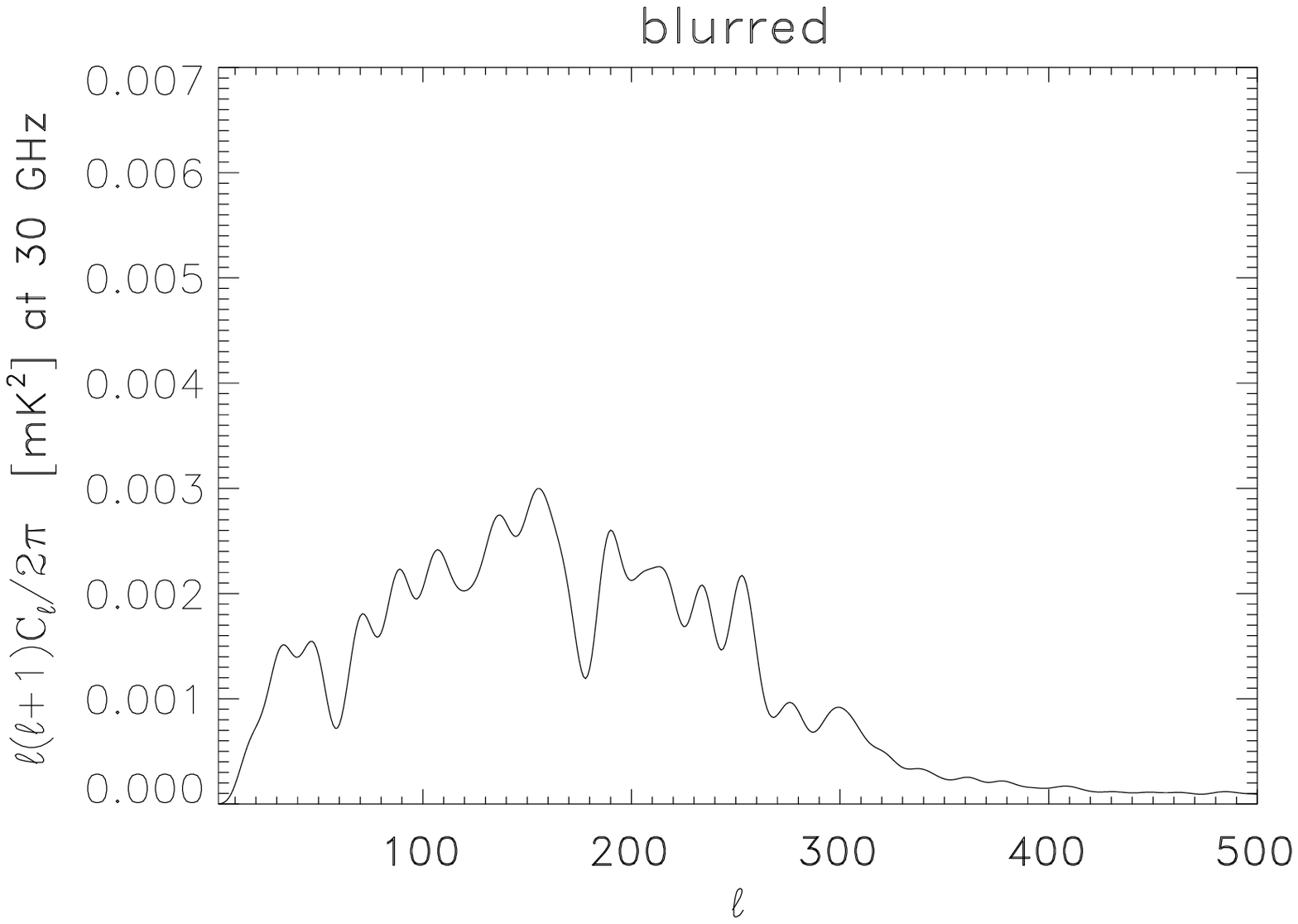}}
        \resizebox{\hsize}{!}{\includegraphics{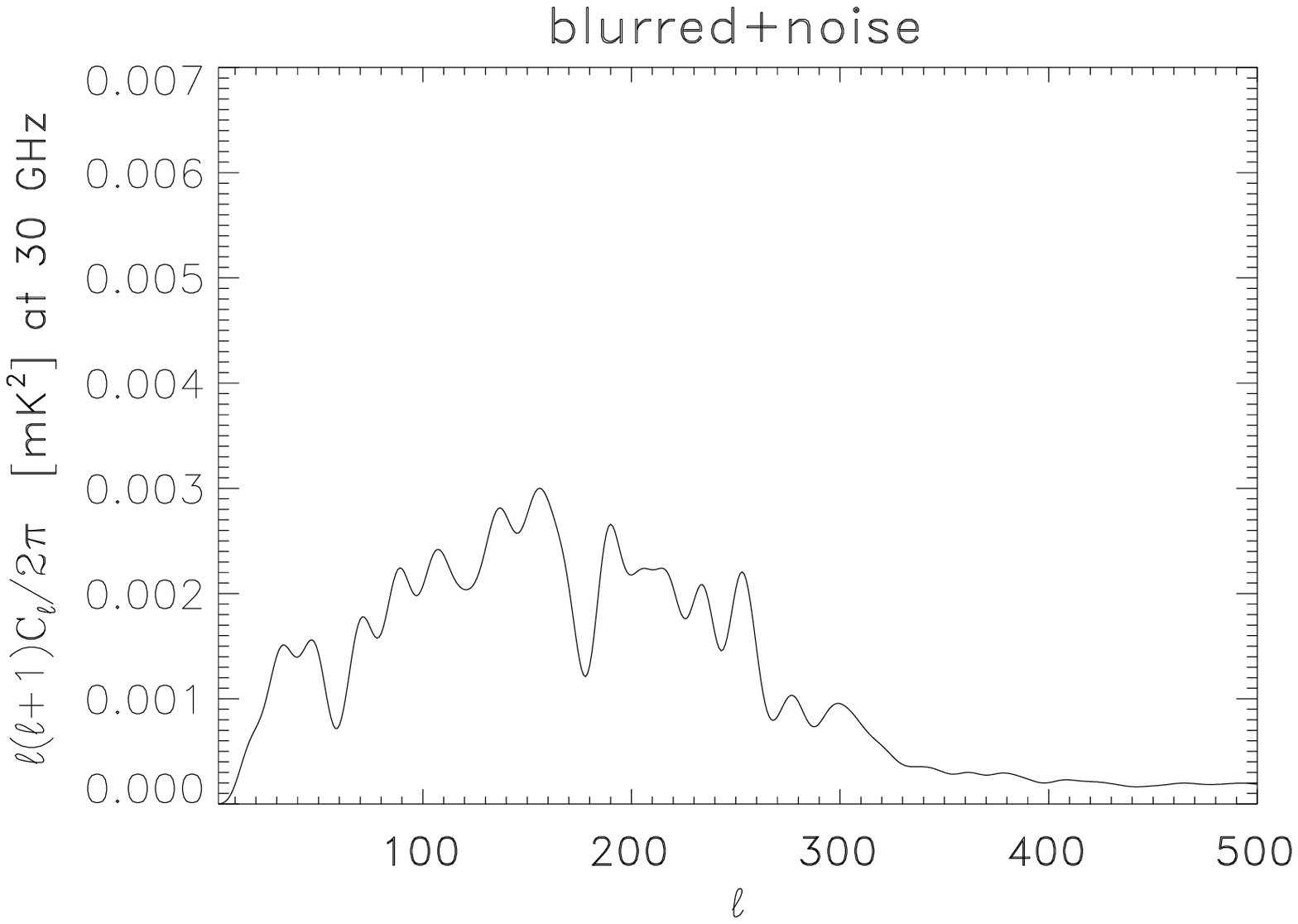}\includegraphics{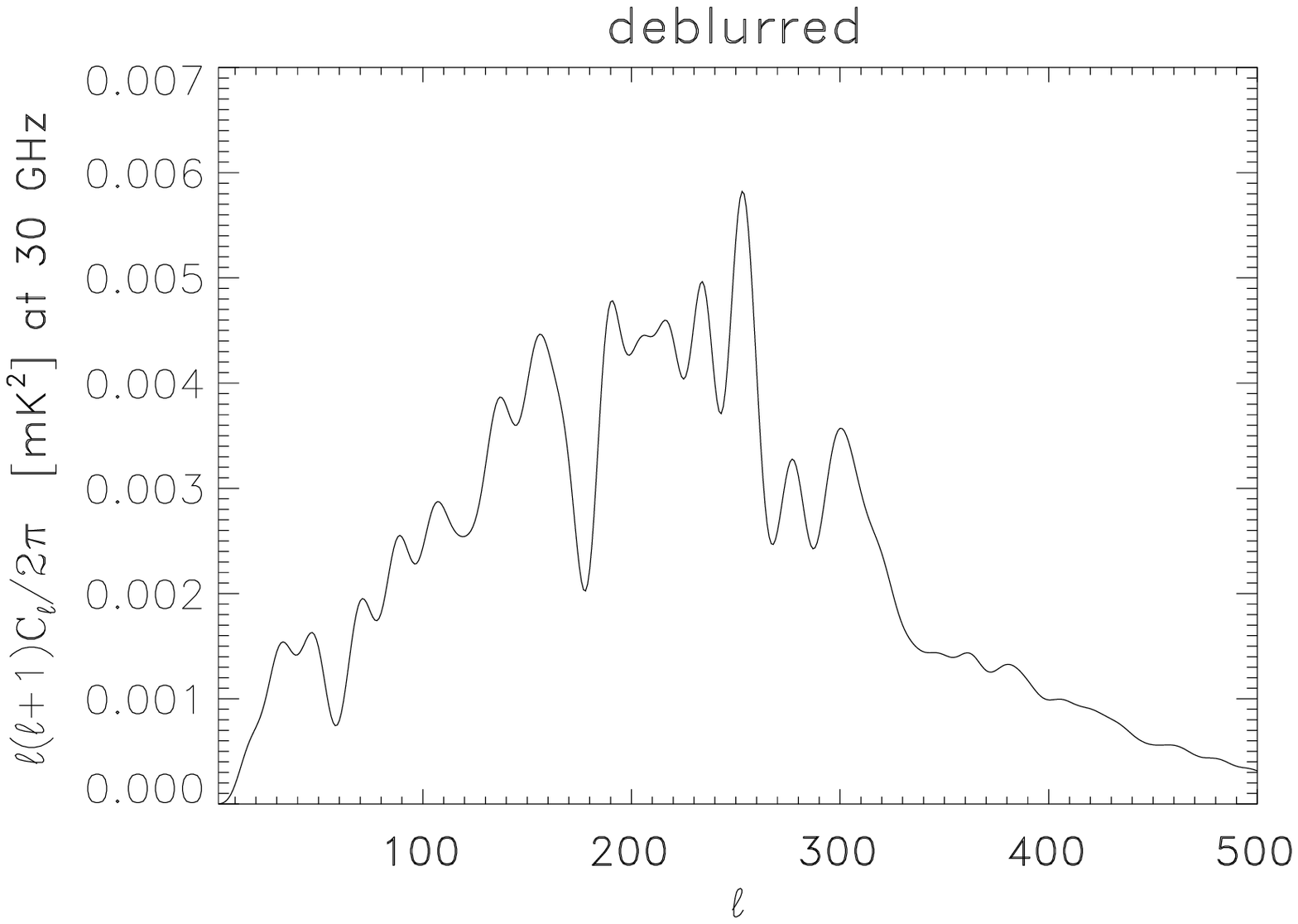}}
        \caption{Angular power spectrum in different steps of the analysis at 30 GHz.
         In the top left panel the plot also shows the shape of the instrumental PSF
         in harmonic space (dotted line).}
        \label{fig:ps30}
\end{figure}
\begin{figure}
        \resizebox{\hsize}{!}{\includegraphics{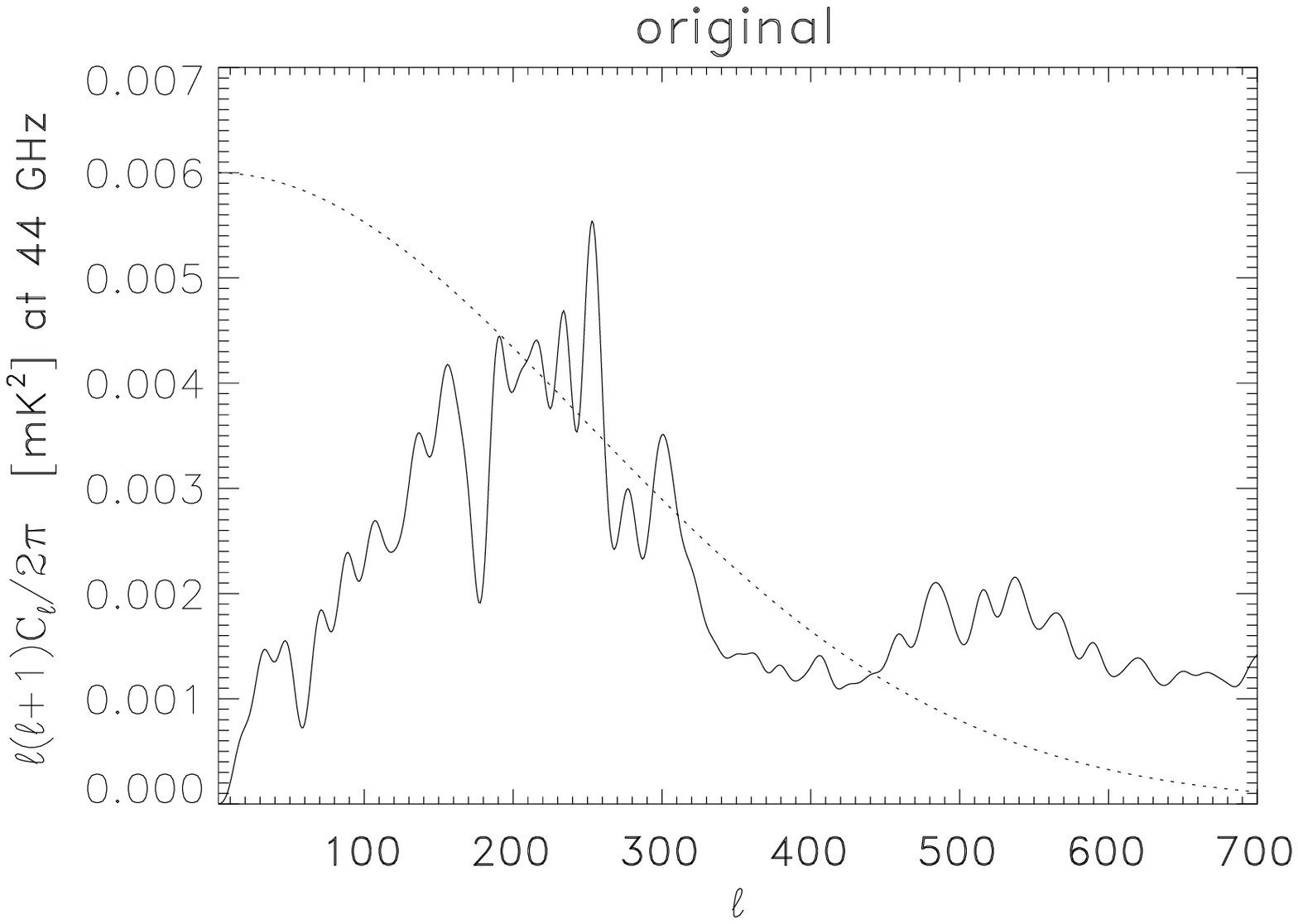}\includegraphics{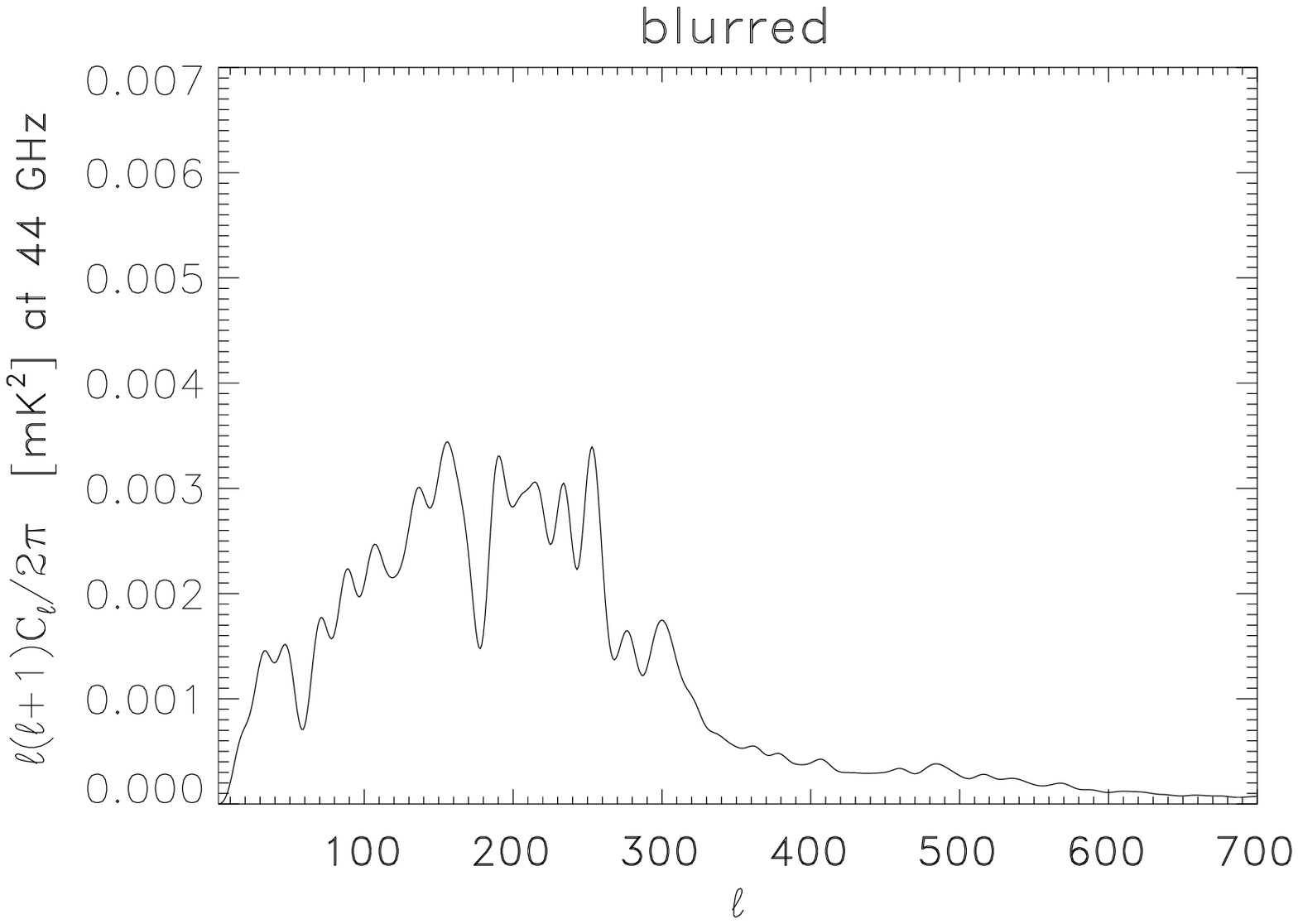}}
        \resizebox{\hsize}{!}{\includegraphics{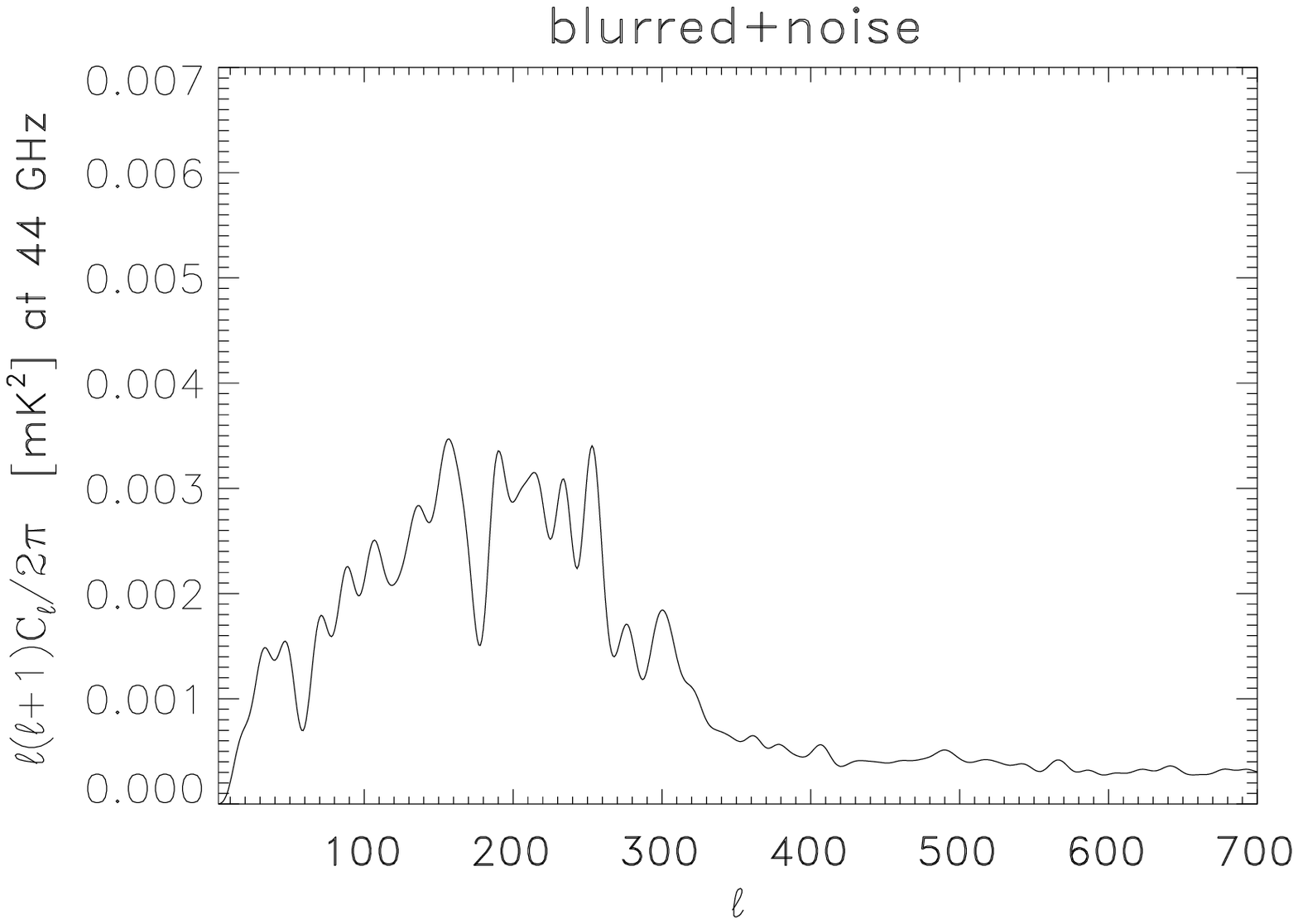}\includegraphics{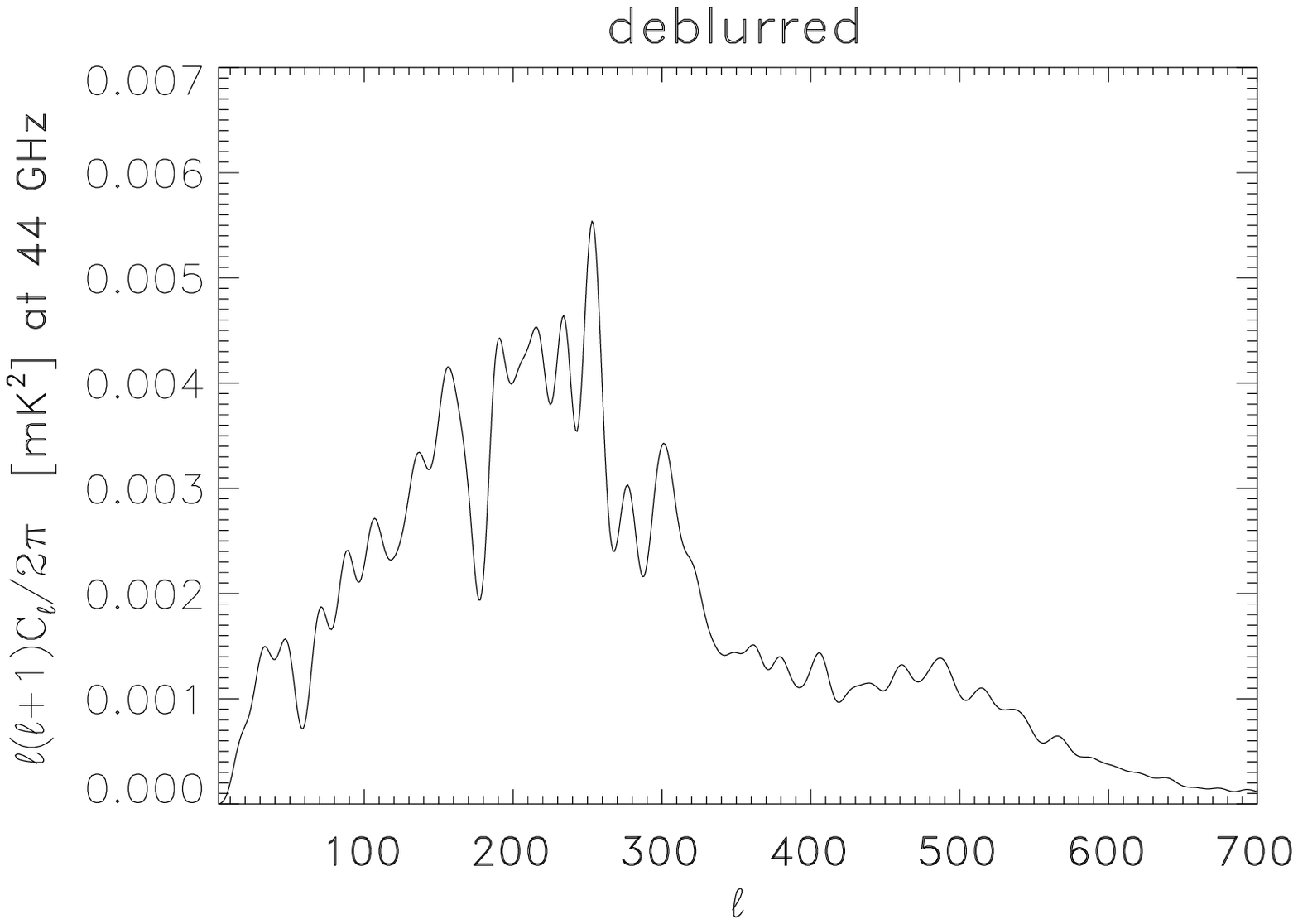}}
        \caption{Same as Fig.~\ref{fig:ps30} but at 44 GHz.}
        \label{fig:ps44}
\end{figure}
\begin{figure}
        \resizebox{\hsize}{!}{\includegraphics{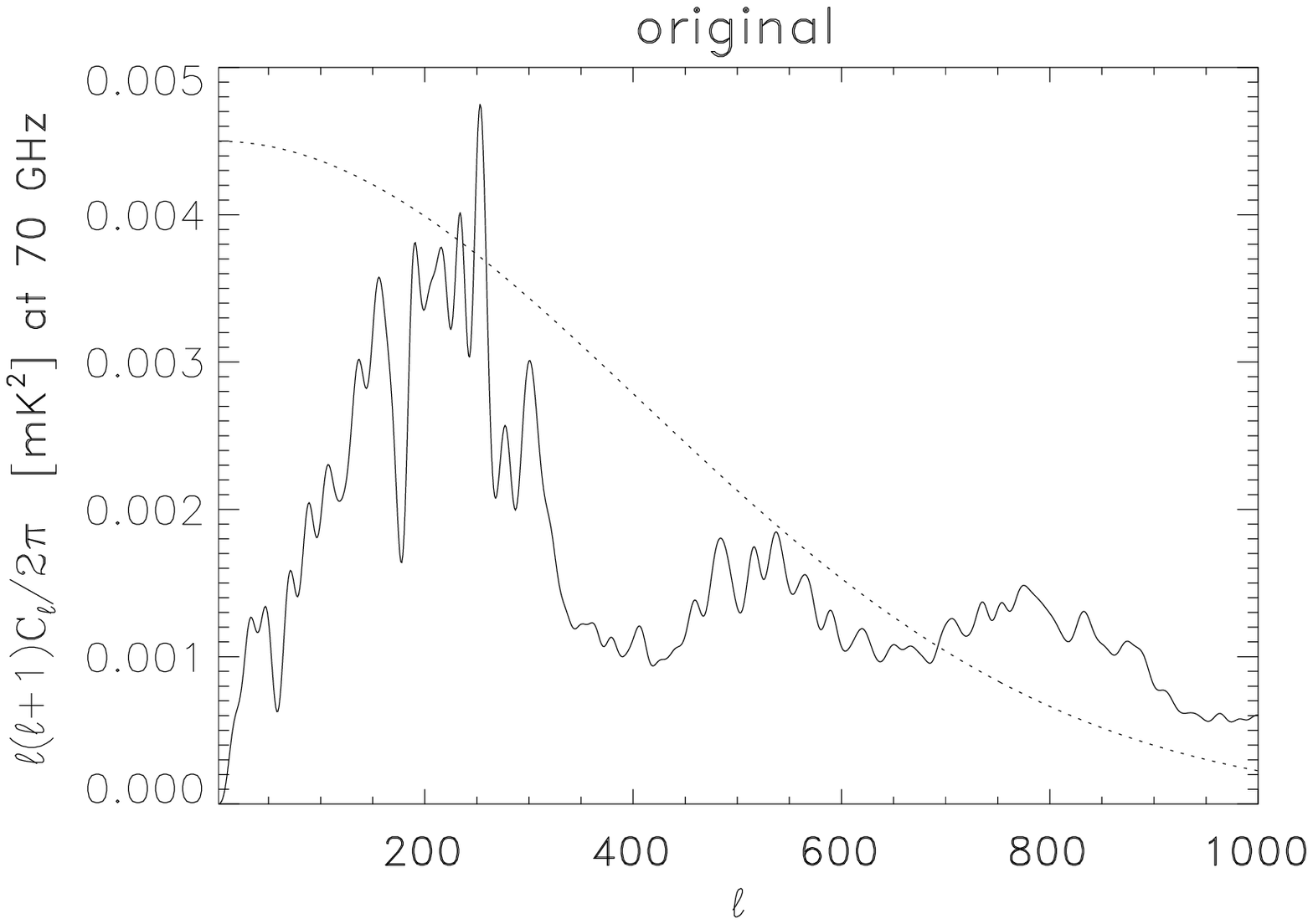}\includegraphics{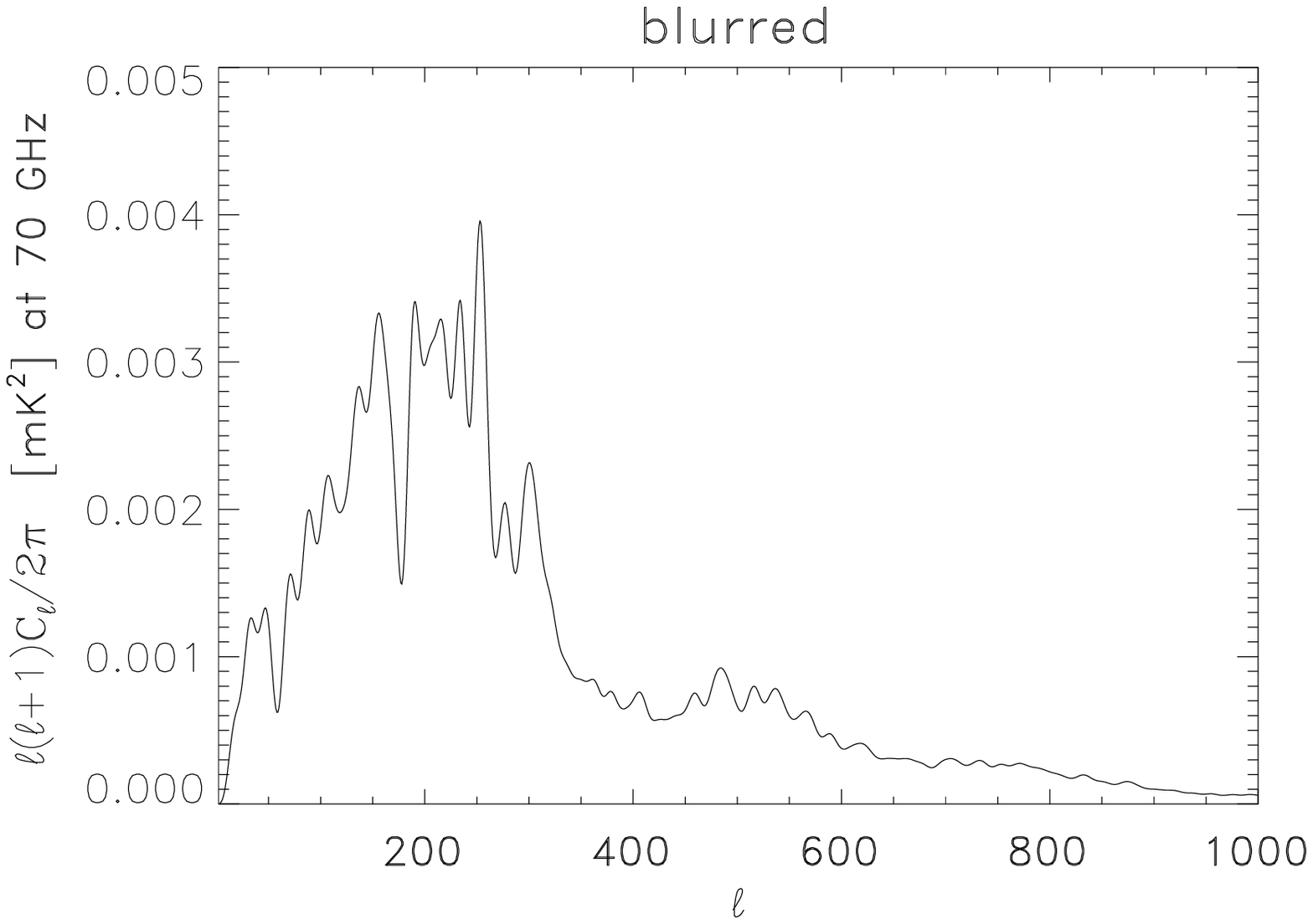}}
        \resizebox{\hsize}{!}{\includegraphics{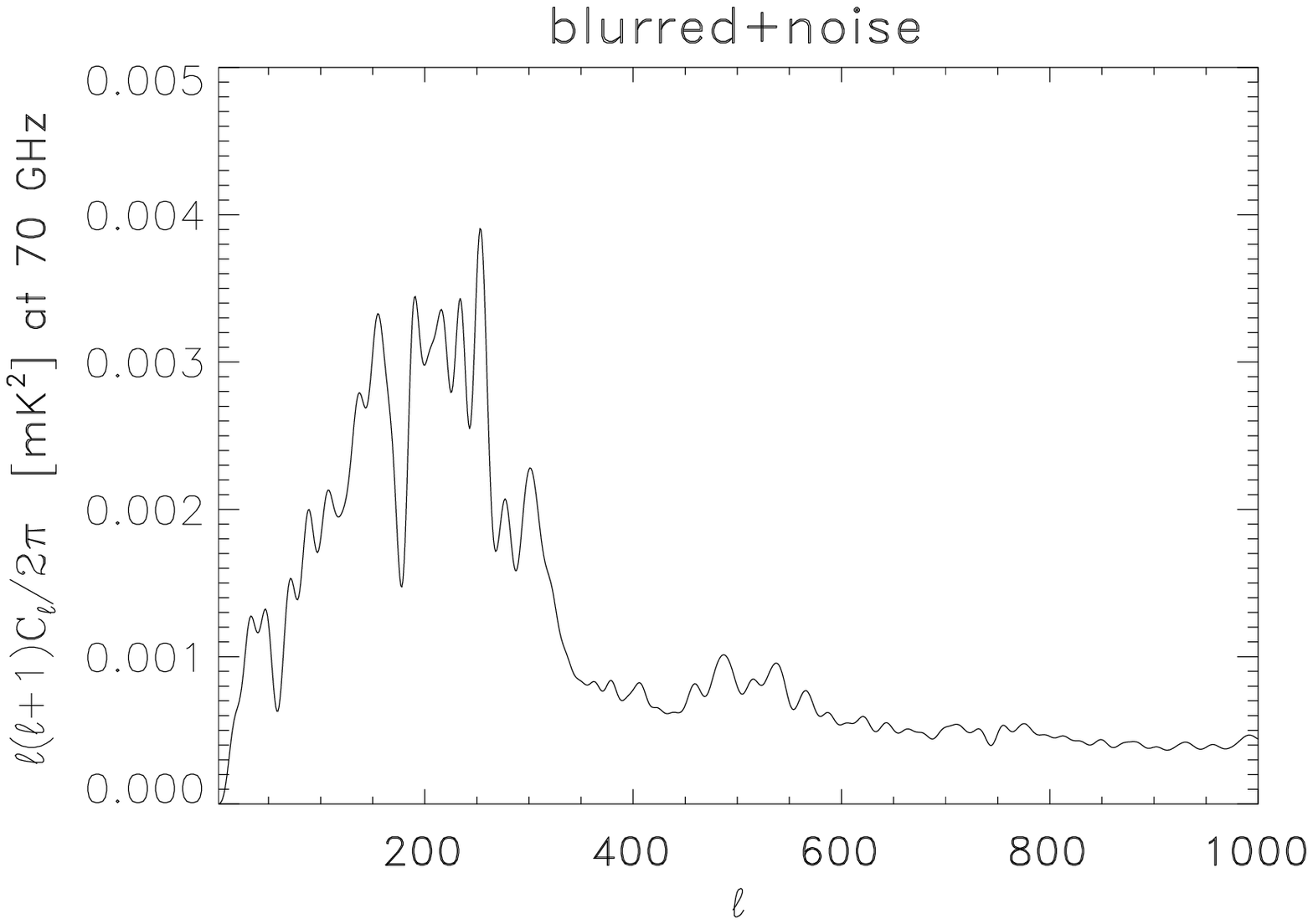}\includegraphics{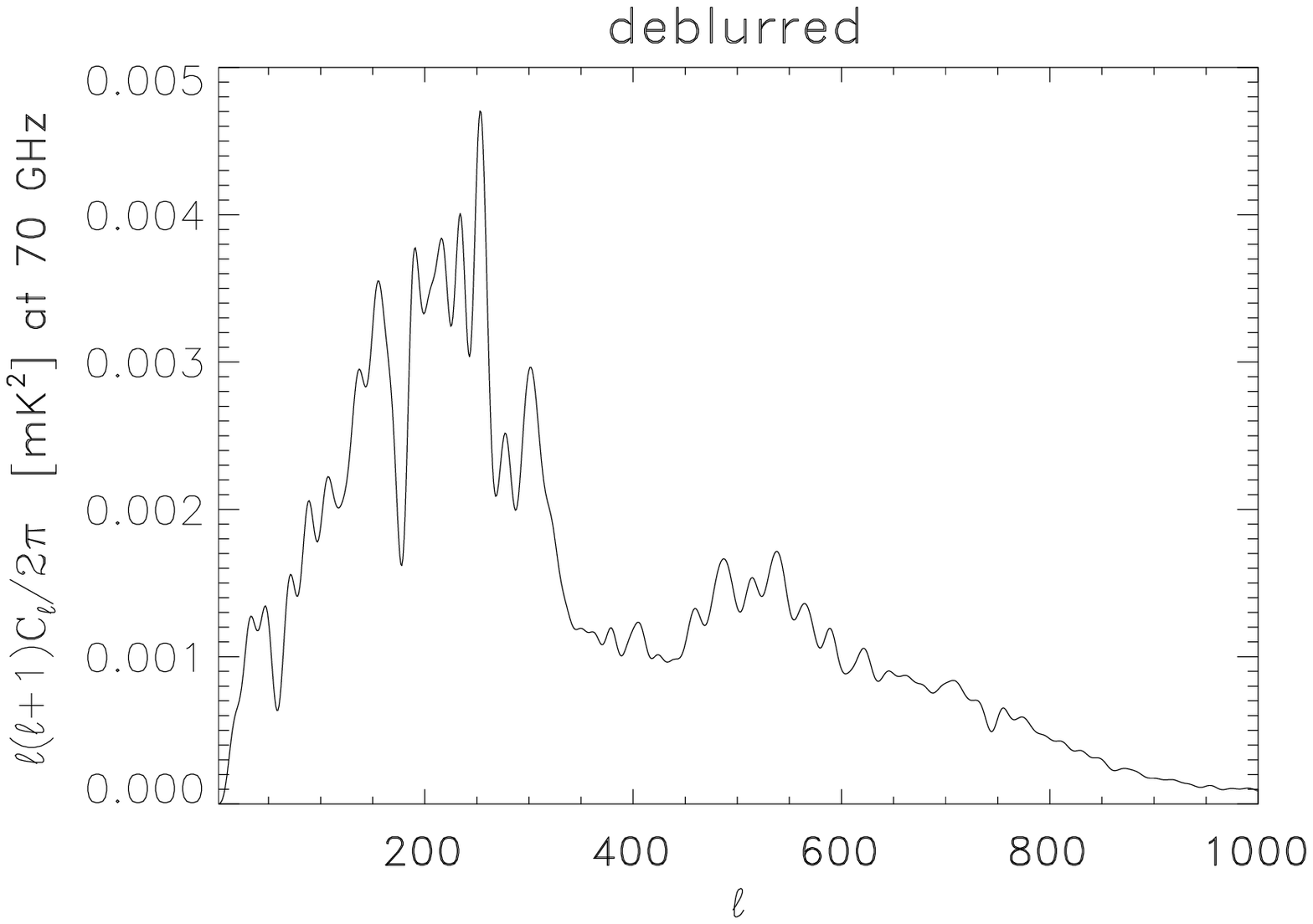}}
        \caption{Same as Fig.~\ref{fig:ps30} but at 70 GHz.}
        \label{fig:ps70}
\end{figure}
\begin{figure}
        \resizebox{\hsize}{!}{\includegraphics{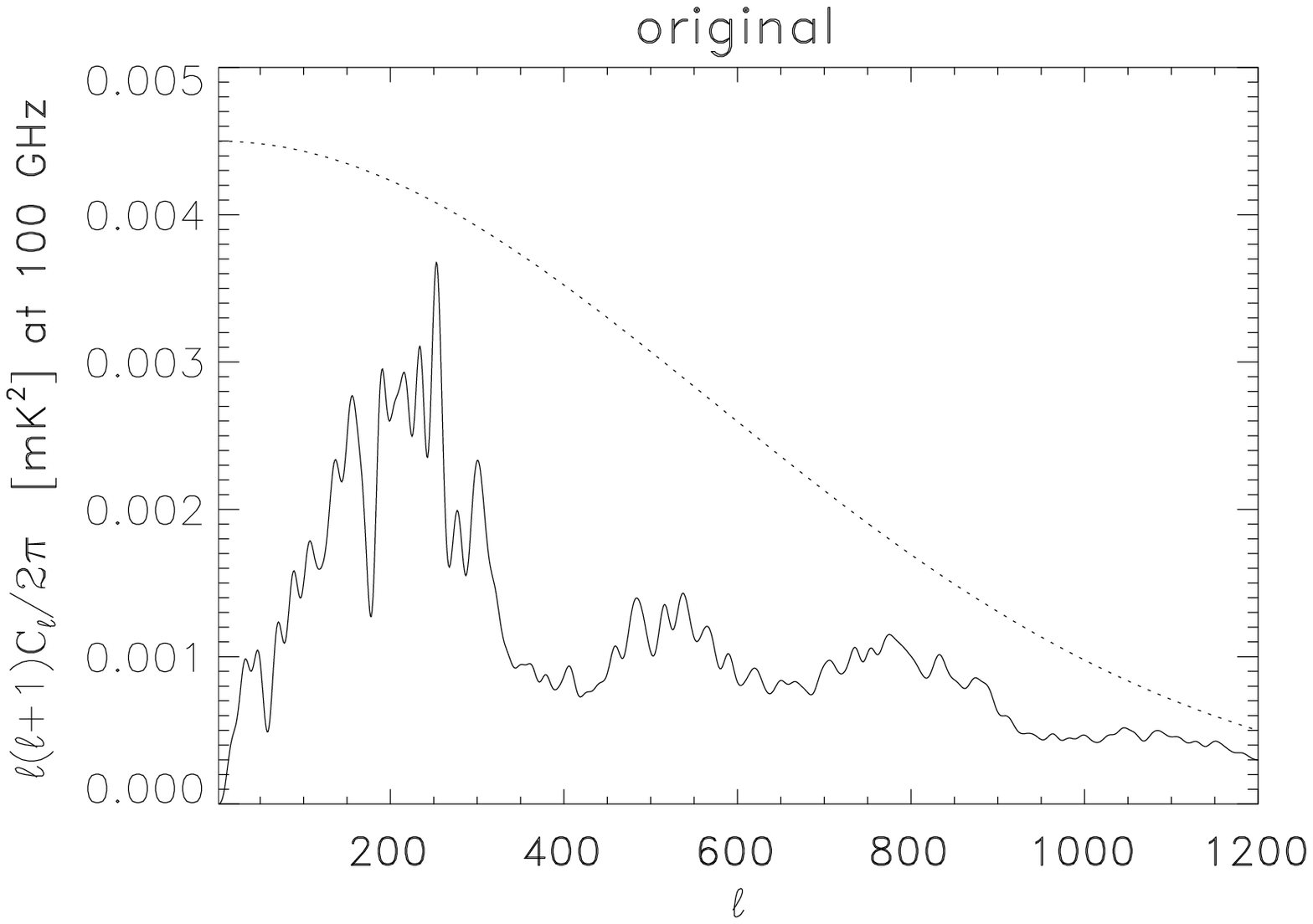}\includegraphics{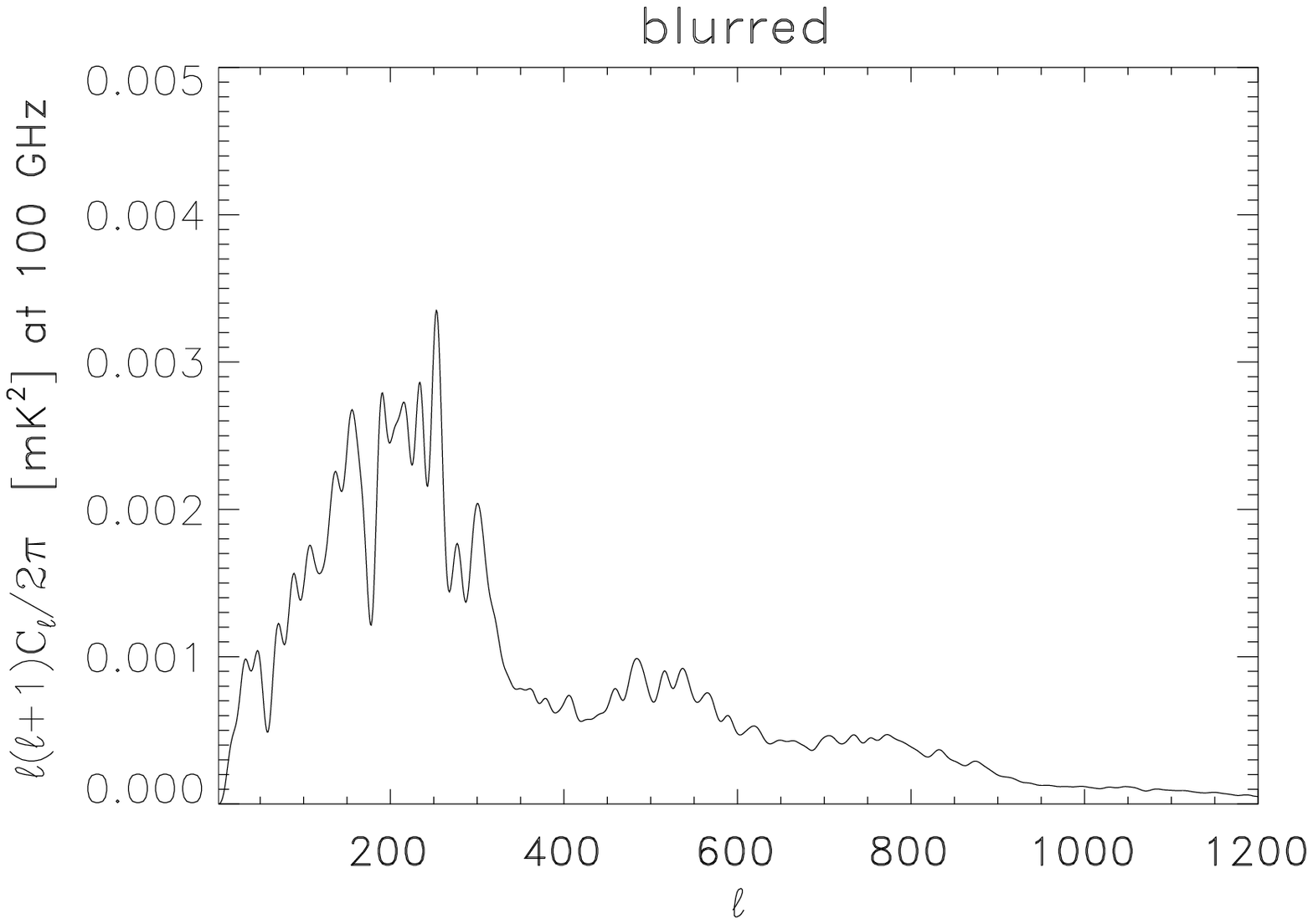}}
        \resizebox{\hsize}{!}{\includegraphics{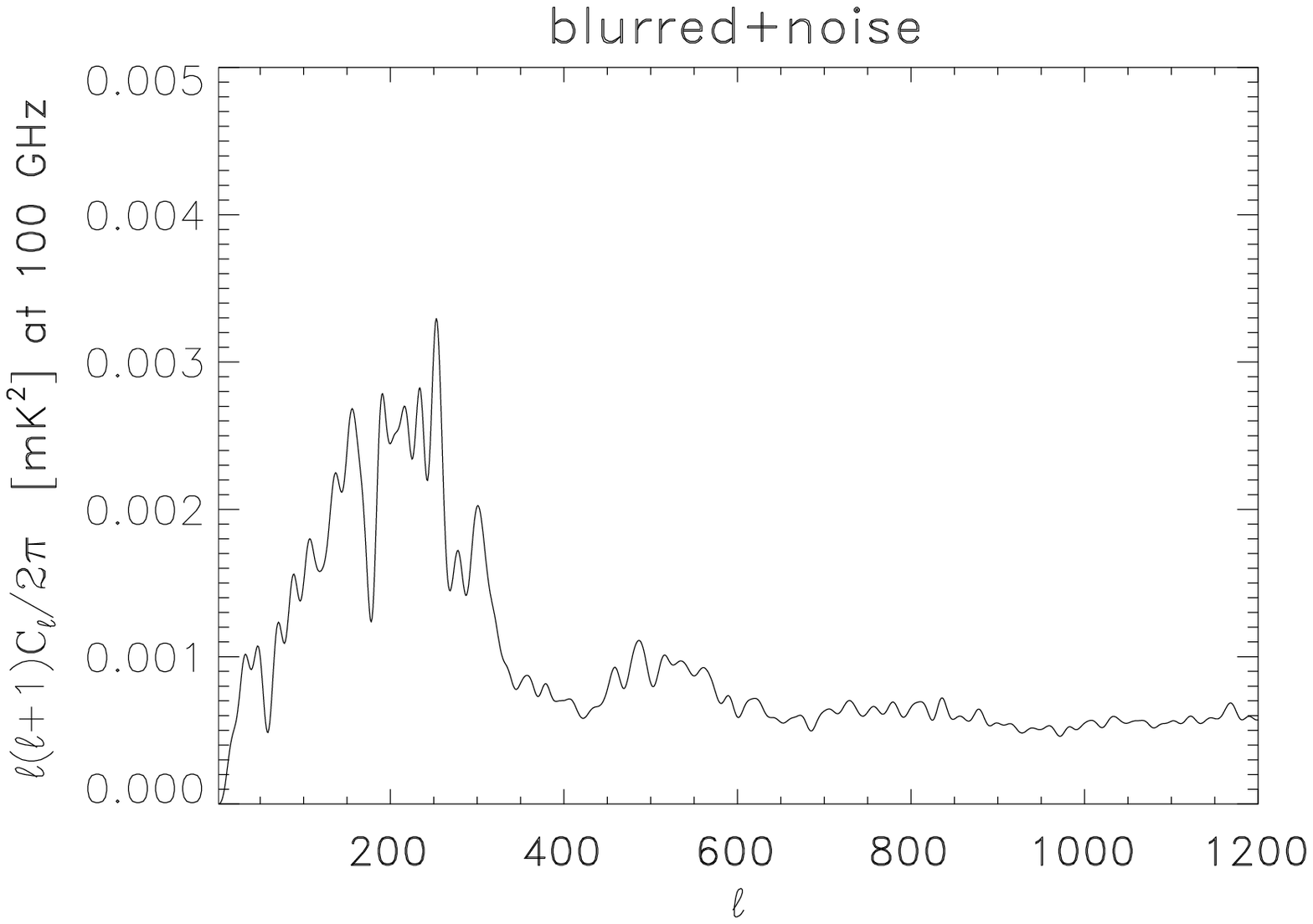}\includegraphics{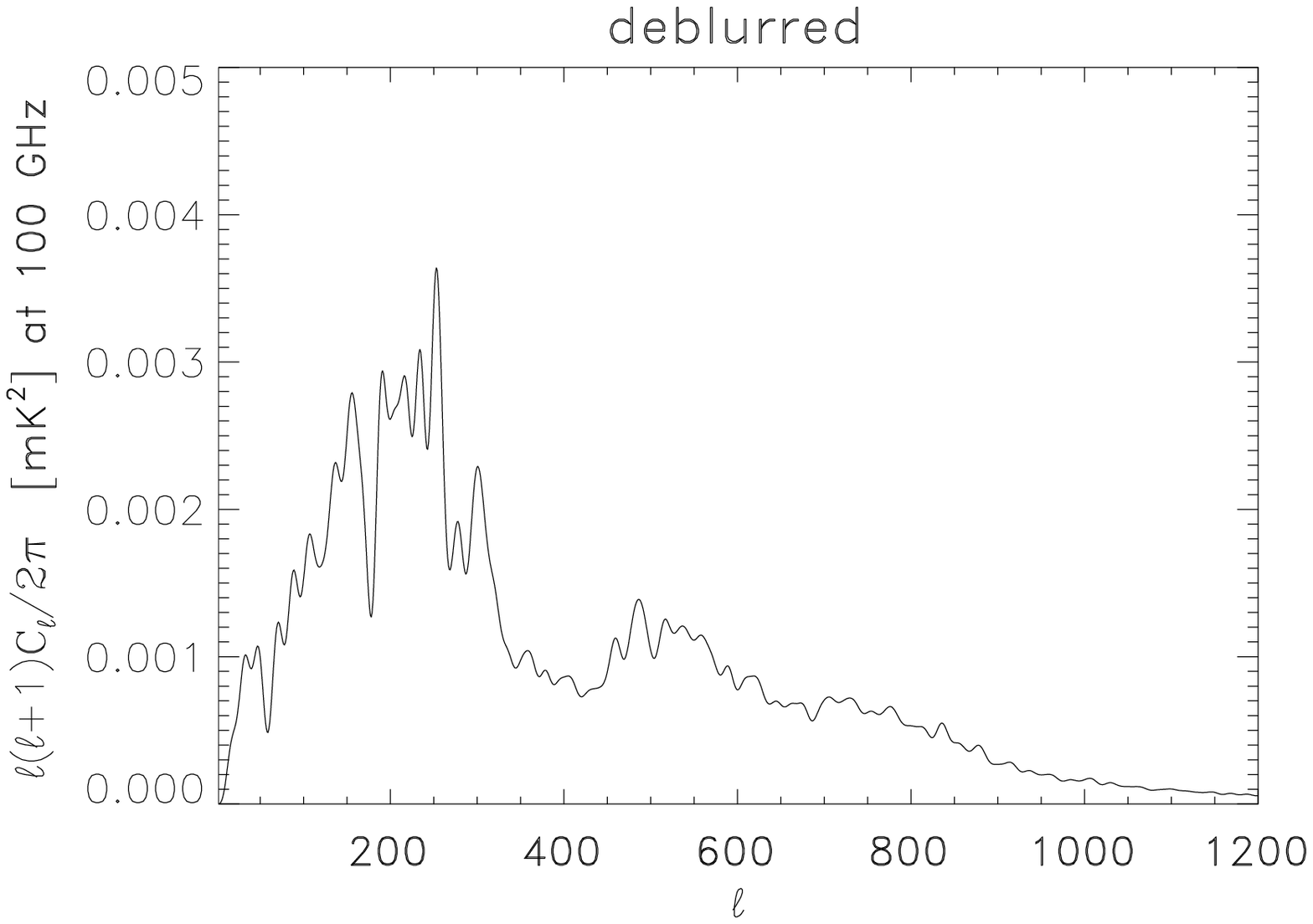}}
        \caption{Same as Fig.~\ref{fig:ps30} but at 100 GHz.}
        \label{fig:ps100}
\end{figure}

As shown in the top panels of
Figs.~\ref{fig:ps30}-\ref{fig:ps100}, the $C_{\ell}$
are increasingly affected as the
frequency decreases; the increasing tail at high multipoles is the effect
of the instrument noise. In all four cases the deblurring procedure has
two main effects. First, it restores amplitude and shape of the part
of the spectrum not dominated by effects of instrument noise and PSF.
Second, as expected from Figs.~\ref{fig:cross1}-\ref{fig:cross2},
it reconstructs part of the power where the PSF causes a major
degradation. In the 30 GHz case, the reconstruction is better in the
multipole range $300\le\ell\le 400$. Similarly, at $44 ~{\rm GHz}$, anisotropy
power is reconstructed up to $\ell\simeq 500$, and up to $\ell\simeq 700$ at
$70$ GHz. Finally, at $100$ GHz part of the original power is recovered up to
$\ell \simeq 800$.

Repeting the same simulations with PSFs of elliptical symmetry gives similar results.

In Summary, in all four channels the deblurring procedure
was effective in recovering spectral properties of the maps.
Although these results have been obtained under the restrictive assumptions of
white noise, the characteristics of the deblurring procedure shown above are
encouraging and deserve more attention and development in future
work.

A last comment regarding component separation. In principle, the separation can be obtained
if maps of the same regions are available at different frequencies
\citep{mai02, hob98}. In practice, this problem is technically difficult and is
far from being solved.
Its treatment is beyond the scope of this paper, but the important point is  that maps
must have the same spatial resolution in order
to carry out the separation. This does not imply, however, that maps have to
be deblurred to a common resolution. One can also achieve a common resolution by blurring
(smoothing). That is, we take Eq.~(\ref{eq:modeld}) and apply a frequency dependent smoothing
operator $\Smat$ to both sides
\begin{equation}
{\gbt} = \Smat\gb = \Smat\fb + \Smat\zb= {\fbt} + {\zbt}.
\end{equation}
Component separation can be done using the different estimates of $\fbt$ obtained by smoothing the
maps of different frequencies.
But without prior information on $\fbt$ the best linear estimates are simply the maps $\gbt$.
To include some prior information we can use smoothness constraints via Tikhonov regularization. The
problem thus reduces to the case we have considered before with the difference that this time
the noise $\zbt$ is correlated due to the effect of $\Smat$. Depending on the noise structure,
one may be able to model correlated noise
in Tikhonov regularization using what is known as {\it mixed effects models}, some examples can be found in
\citet{ping}\footnote{http://www.stat.purdue.edu/$\sim$chong} and  \citet{robinson}.
If the noise is stationary, one can deconvolve in the frequency domain but again, when the signal
spectrum is unknown we may want to regularize the problem using Tikhonov in the frequency domain.

\section{An example of deblurring effects on non-Gaussianity} \label{sec:nongauss}

An important point in CMB research is the detection
of non-Gaussianity. It is therefore of interest to check the
effects of the deblurring procedure on the non-Gaussian characteristics
of a map.

Experimental results have not yet found any evidence of non-Gaussianity in CMB maps. It is
even difficult to conduct realistic simulations of non-Gaussian CMB maps since it is not clear
what type of non-Gaussian behavior one should look for.
As an example, we consider the effects of deblurring on the marginal distribution
of a particular homogeneous non-Gaussian random field whose marginal distribution is slightly different
from a Gaussian (skewness $\simeq 0.22$
and kurtosis $\simeq -0.11$ -- see Fig.~\ref{fig:pdf}). The field has the same autocorrelation function,
PSF (with circular symmetry) and
$S/N$ of the Gaussian fields simulated in Sect.~\ref{sec:numerical}.
Realizations of this field are simulated via the method
presented in \cite{vio01}.
Skewness and kurtosis coefficients (normalized to be zero for a Gaussian
distribution) are calculated for each simulation.

Table~\ref{tab:nongauss} presents the average change, from the original values, in skewness and kurtosis
over 100 simulations.
As expected, noise and blurring have the effect of ``{Gaussianizing}'' the random fields
(i.e., the estimated skewness and kurtosis are closer to zero than the original values), with obvious consequences
on the response of any non-Gaussianity test. Deblurring noticeably improves estimates.

These results, together with the power spectrum reconstruction discussed before, provide a further indication
of the usefulness of deblurring in the analysis of CMB maps.

\begin{table*}
\begin{center}
\begin{tabular}{l c c c}
\hline
\hline
map & $\Delta$ skewness  & $\Delta$ kurtosis \\
\hline
30 GHz, blurred & $-0.052 \pm  0.058$  & $+0.022 \pm 0.123$ \\
30 GHz, blurred noisy & $-0.098 \pm 0.029$ & $+0.055 \pm 0.054$ \\
30 GHz, deblurred & $-0.035 \pm 0.024$ &  $+0.032 \pm 0.054$ \\
\hline
44 GHz, blurred & $-0.038 \pm 0.040$ & $+0.021 \pm 0.087$ \\
44 GHz, blurred noisy & $-0.088 \pm 0.017$ &  $+0.054 \pm 0.031$ \\
44 GHz, deblurred & $-0.030 \pm 0.019$ & $+0.025 \pm 0.042$ \\
\hline
70 GHz, blurred & $-0.024 \pm 0.024$  & $+ 0.016 \pm 0.053$ \\
70 GHz, blurred noisy & $-0.078 \pm 0.008$ & $+0.051 \pm 0.018$ \\
70 GHz, deblurred & $-0.024 \pm 0.014$ & $+0.020 \pm 0.032$ \\
\hline
100 GHz, blurred & $-0.018 \pm 0.017$ & $+0.012 \pm 0.038$ \\
100 GHz, blurred noisy & $-0.074 \pm 0.008$ & $+0.049 \pm 0.020$ \\
100 GHz, deblurred & $-0.022 \pm 0.013$ & $+0.018 \pm 0.027$ \\
\hline
\hline
\end{tabular}
\caption{Mean and standard deviation of the difference between the skewness and kurtosis
coefficients of the blurred, blurred+noise, and deblurred
non-Gaussian maps with respect to corresponding values (respectively $\approx 0.22$ and
$ \approx -0.11$) of the original map (see text). The results are based on 100 simulations
with $S/N=2$.}
\label{tab:nongauss}
\end{center}
\end{table*}

\section{An example of deblurring with spatially varying PSF} \label{sec:variant}

Although the PSF of {\it PLANCK}'s instrument is designed to be spatially invariant,
this condition may change after launch, and there may be other experiments that require
more general deblurring methods that can be used with spatially varying PSF.

In principle, a real space approach to restoring an image degraded by
a spatially variant PSF can be developed, but in practice the methods
are quite difficult to implement.  Indeed, efficient
implementations are currently available for iterative methods \citep{NaOL97}, but
direct algorithms similar to those presented in previous sections
for invariant PSFs have not been developed yet. We are presently working on this problem.
To illustrate the importance of pursuing this work,
we present some simulation results that use an iterative
method to restore an image blurred by a spatially variant PSF.

As claimed in Sect.~\ref{sec:tikmap}, one of the most serious difficulties
with iterative methods is in the choice of
an appropriate stopping criteria.  Though methods such as the
discrepancy principle, L-curve and GCV can be used, our success with
these approaches on CMB maps has been marginal.
Without an efficient method for choosing a regularization parameter,
or for determining an appropriate stopping criteria, iterative
methods may not be the ideal choice for CMB maps. However, because
matrix-vector multiplications with $\Hb$ and $\Hb^T$ (the most expensive part
in the iteration procedure) can be done efficiently for spatially varying PSF
\citep{NaOL97}, iterative
methods do give us a means of determining if better restorations can be obtained
with a spatially varying model.

To construct a test example, we have used a PSF that changes linearly across the map
as shown in Fig.~\ref{fig:psf}. Deblurring was done using a standard conjugate gradient
iterative method \citep{Hansen97}, stopping the iteration
when the computed restoration is closest, in a mean square sense,
to the true image.  The matrix $\Hb$ modelling the spatially varying PSF
is constructed by assuming that it is approximately spatially invariant in small regions.
If the corresponding, spatially invariant matrix defined by the PSF in the $i$th region is $\Hb_i$,
then the spatially varying matrix $\Hb$ is defined as
\[
  \Hb = \sum_{i=1}^m \Db_i \Hb_i\,,
\]
where the matrices $\Db_i$ are nonnegative diagonal and $\sum \Db_i = \Ib$, where $\Ib$ is the identity matrix.
Note that using this $\Hb$ is equivalent to interpolation to match the individual PSFs across the different
patches.  For example, for piecewise constant interpolation the $j$th diagonal entry of $D_i$ is 1 if the $j$th
pixel is in region $i$, and 0 otherwise.
Efficient matrix vector multiplications with $\Hb$ and $\Hb^T$ exploit the sparsity of $\Db_i$ and the spatially
invariant structure of $\Hb_i$.  The implementation details are tedious to describe; we only mention here that
the basic idea is related to overlap-add and overlap-save convolution methods, and refer
the interested reader to \citep{NaOL97} for a description of the algorithms, and to \citep{LeNaPe02} for a
Matlab implementation.

The matrix $\Hb$ in our simulations was constructed through piecewise constant interpolation of
121 PSFs uniformly distributed across the map.
The computed spatially varying restoration and the corresponding spatially invariant
restoration  are shown in Fig.~{\ref{fig:variant}}. We have used a signal to noise ratio of $S/N=4$
because the iterative procedure is more sensitive to noise than direct methods.
Despite these limitations, Fig.~\ref{fig:psf} shows the advantage of deblurring with a spatially
varying PSF.

\section{Discussion and conclusions} \label{sec:conclusions}

We have considered Tikhonov regularization for deblurring CMB maps in real space.
Although more demanding from the computational point of view, this approach permits
the development of algorithms that are more flexible and robust than those based on frequency-space
methods. Furthermore, as shown in Fig.~\ref{fig:time}, the computational cost
can be significantly reduced by carefully implementing the algorithms to take advantage of
the characteristic structure of the matrices involved.

We have applied the Tikhonov methodology to
simulated skies at typical CMB frequencies.
We considered test signals with known statistics,
as well as realistic simulations of the CMB sky
contaminated by noise whose rms is that expected for the low frequency
instrument aboard the {\it PLANCK} satellite.
This case is particularly interesting for
application of a deblurring procedure, as the
instrument observes the sky at 30, 44, 70 and 100 GHz
with very different PSFs of resolution $33$, $22$, $14$, $10$
arcminutes.

We analysed the effects of the deblurring
procedure by studying different characteristics of the restored image.
Contour plots of the two-dimensional cross-correlation
functions of the original and the deblurred maps show that
the algorithm effectively improves the resolution, especially that of the
worst resolution channels at 30 and 44 GHz. The same effect can be seen
in the multipole coefficients of the angular power spectra; the original power on angular scales hidden by the
instrument's PSF is recovered on a significant range of multipoles.
We also performed an example of skewness
and kurtosis recovery by the deblurring procedure.
Instrumental noise and PSFs generally have the effect of
pushing skewness and kurtosis toward their Gaussian values; deblurring brings these values closer to
their true non-Gaussian ones.

On the basis of these promising results, we
plan to develop deblurring algorithms based on the methods we have presented.
Since
Satellite CMB experiments are able to perform
all-sky observations, a robust
deblurring procedure which is able to work on
the whole sphere is conceivable.
We also plan to test map based component
separation techniques, requiring multifrequency data of
the same resolution, on deblurred maps.

\begin{figure}
\resizebox{\hsize}{!}{\includegraphics{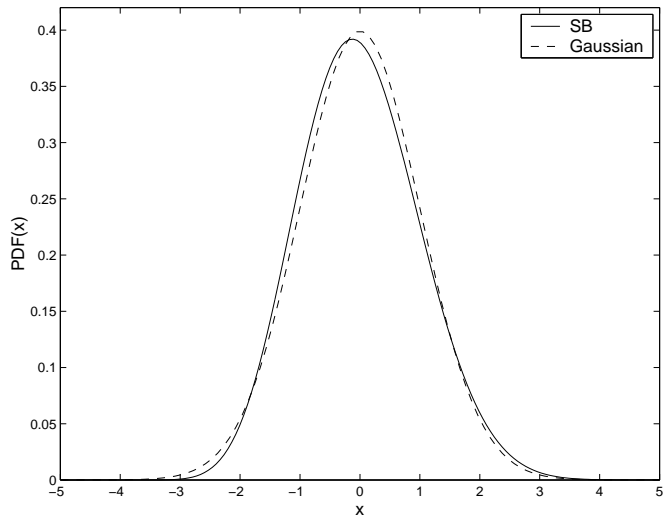}}
\caption{Probability density function used in the simulation
of the non-Gaussian random fields (see text). It corresponds to a SB
distribution belonging to the Johnson's
parametric family \citep[see][]{vio94}.}
\label{fig:pdf}
\end{figure}

\begin{figure}
        \resizebox{\hsize}{!}{\includegraphics{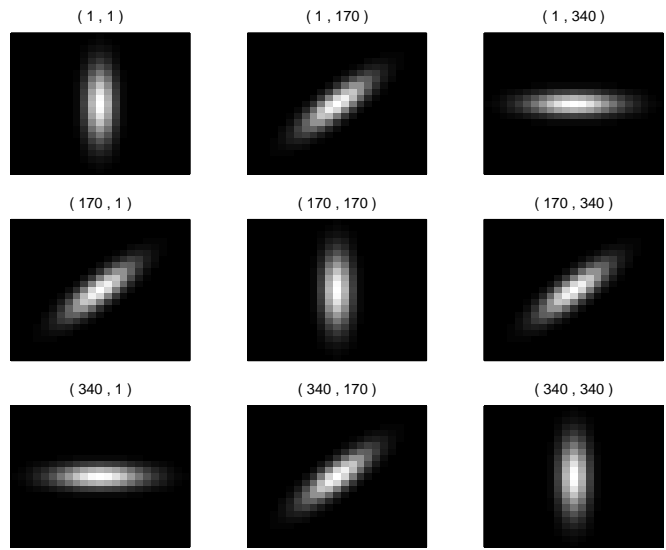}}
        \caption{Some of the PSF's used in the experiment of the variant PSF described in the text. The couple
        of numbers in the header of each panel provides the coordinates of the pixels to which the displayed PSF
        corresponds. Each PSF is given by a two-dimensional Gaussian function with ${\rm FWHM}=33 ~{\rm arcmin}$
        along the major axis and ${\rm FWHM}=10 ~{\rm arcmin}$ along the minor axis. The size of each panel is
        $23 \times 23$ pixels.} \label{fig:psf}
\end{figure}

\begin{figure}
        \resizebox{\hsize}{!}{\includegraphics{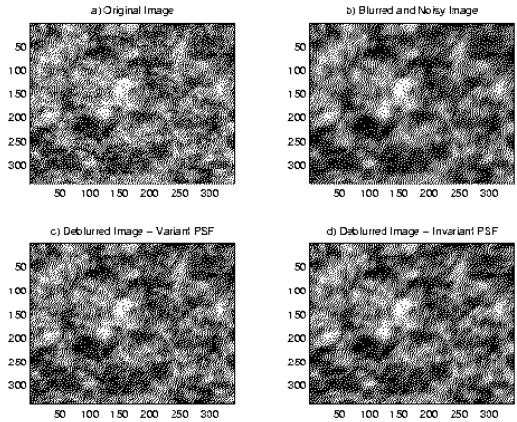}}
        \caption{
        a) original map; b) map obtained by blurring the map in the previous panel with a PSF
        changing uniformly its orientation across the frame. Some of the PSF's are
        shown in Fig.~(\ref{fig:psf}). White noise has been added to the map ($S/N=4$); c)
        map obtained by deblurring with the spatial variant method explained in the text. A set of $121$ PSF's
        uniformly distributed across the map have been used; d) map obtained by deblurring with a spatially invariant
          PSF. The PSF used is that shown in uppermost-left panel in
          Fig.~(\ref{fig:psf}). The standard deviation of the difference between the deblurred and the true
        maps is $0.41$ for the spatially varying PSF and $0.43$ for the spatially invariant one.}
        \label{fig:variant}
\end{figure}
\begin{figure}
        \resizebox{\hsize}{!}{\includegraphics{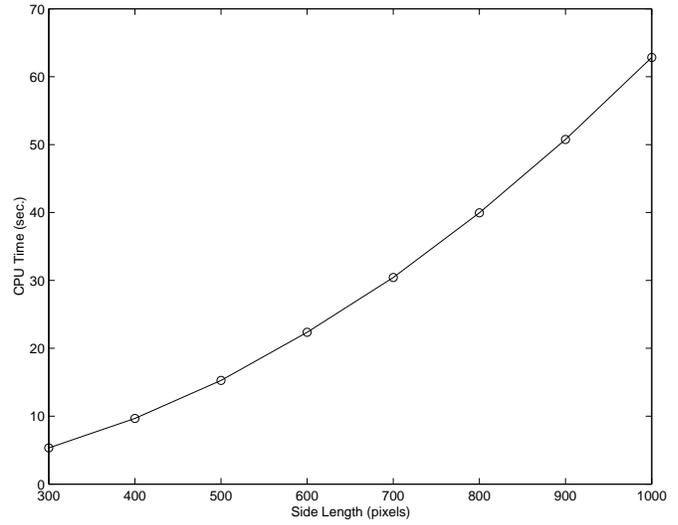}}
        \caption{CPU time (sec.) required by the full Tikhonov
        deblurring procedure of $N\times N$ maps as a function of $N$.
        The PSF is the one used for the simulations of CMB maps at $30~{\rm GHz}$ (circularly symmetric
        Gaussian FWHM $\approx 9.5$ pixels). Similar results are obtained for elliptic PSF with
        ${\rm FWHM}=9.5$ and ${\rm FWHM}=2.8$ pixels along the major and minor axes, respectively, and
        whose orientations are parallel to the sides of the map. Experiments have been conducted with Matlab 6
        on a {\it Pentium IV} - $1500~{\rm GHz}$ processor in a {\it Windows 2000} operating system.}
\label{fig:time}
\end{figure}

\appendix
\section{Automatic choice of the regularization parameter} \label{sec:gcv}

We briefly describe some methods to estimate the smoothing parameter
$\lambda$, which is an essential  ingredient in Tikhonov regularization.
For further discussions of this topic see \cite{Hansen97,ten01,vog02}.

Since the data $\gb$ are noisy observations of $\Hb\fb$,
it seems reasonable to choose a value of $\lambda$ that minimizes the predictive mean square error (${\rm PMSE}$),
\begin{equation}
{\rm PMSE}(\lambda) =  \frac{1}{n}\,\|\,\Hb \fb - \Hb \fb_\lambda\,\|^2.
\label{pmse}
\end{equation}
But since $\fb$ is unknown, we minimize instead a cross-validation (CV) estimate of (\ref{pmse}) obtained
by plugging in (\ref{pmse})
the data as a proxy for $\Hb \fb$ and a leave-one-out estimate for  $\Hb \fb_\lambda$. The result can be written
as
\begin{equation}
{\rm CV}(\,\lambda\,)  = \frac{1}{n}\,\sum_{i=1}^n \left( \frac{g_i - \widehat{g}_{i,\lambda}}
{1-\Hcb_{ii}(\lambda)}\right)^2 ,
\end{equation}
where $\Hcb_{ii}(\lambda)$ are the diagonal elements of the
``hat'' matrix (\ref{hat:eq}) that maps $\gb$ into $\gbh_\lambda=\Hcb \fb_\lambda$.
The CV estimate of $\lambda$ is the value that minimizes ${\rm CV}(\lambda)$.

The generalized cross-validation (GCV) function is a ``smoothed'' version
of ${\rm CV}$ in which the diagonal elements of $\Hcb(\lambda)$ are replaced by their
average
\begin{equation}
{\rm GCV}(\lambda) = \frac{ \Vert\, \,\gb - {\gbh}_\lambda\,\,\Vert_2^2/n}
{(\,1-\,{\rm trace} \,\Hcb(\lambda)/n)^2}.
\label{gcveq}
\end{equation}
Unlike cross-validation, ${\rm GCV}$ is invariant under orthogonal transformations of the data.

A slight modification of GCV provides and estimate $\widehat{\sigma}^2$ of the noise
variance $\sigma^2$
\begin{equation}
\widehat{\sigma}^2 = \frac{ \Vert\, \,\gb - {\gbh}_\lambda\,\,\Vert_2^2}
{n-\,{\rm trace} \,\Hcb(\lambda)}.
\label{sigest}
\end{equation}
This is just a normalized residual sum of squares where the effective degrees of freedom is determined
by the trace of $\Hcb(\lambda)$.

The L-curve method \citep{Hansen97} is another way to determine a balance between goodness
of fit and roughness. As $\lambda>0$ increases, the points
\begin{equation}
C(\lambda)=(\,\,\|\,\gb-\Hb\fb_\lambda\,\|,\,\, \|\Lb\,\fb_\lambda\,\|\,\,)
\label{lcurveeq}
\end{equation}
define a convex curve on the plane that in the log-scale resembles the
letter ``L''.
The selection of $\lambda$ corresponds to the ``corner'' value where $C(\lambda)$
has the highest curvature.

In general, estimates of $\lambda$ based on cross-validation are robust to small deviations
from the homogeneous variance and Gaussian assumptions and,
for large samples, converge to the optimal minimizer of the PMSE
\citep{wahba90}. When the ${\rm GCV}$ function is almost flat around its minimum, it may lead to very small
values of $\lambda$ (undersmoothing). A lower limit on $\lambda$ that controls undersmoothing
can be achieved by multiplying the trace term in (\ref{gcveq}) by a constant $k>1$ \citep{fried89,gu02}.

The L-curve has not been
studied as much as GCV but some studies seem to indicate
that L-curve estimates have the same
good properties of GCV and, in addition, may be more robust to correlated
noise \citep{Hansen97}. Note, however, that \cite{vog96} has pointed
out some convergence problems with L-curve estimates.

\section{Efficient implementation of the Tikhonov algorithms}
\label{sec:Implementation}

\subsection{Exploiting the structure of $\Hb$ and $\Lb$}
The efficiency of our approach to computing a minimum of the
${\rm GCV}$ function (\ref{eq:gcv1}), and to solving
the least squares problem (\ref{eq:tikls}), is
based on exploiting structure of the matrices $\Hb$ and $\Lb$.
We assume that $\Lb$ is a structured matrix, such as
given in Eq.~(\ref{eq:secondd}).

Fast algorithms for certain structured matrices arising
in image deblurring are well known.
For example, when using periodic boundary conditions with a
spatially invariant blur, the matrices $\Hb$ and $\Lb$
are BCCB, and therefore have the spectral factorizations
\begin{equation}
\label{eq:Pfactor}
  \Hb = \fmatb^* \Sigmab \fmatb\,, \quad
  \Lb = \fmatb^* \Deltab \fmatb\,.
\end{equation}
Here, $\fmatb = \Fb_N \otimes \Fb_M$ with ``$\otimes$'' the
Kronecker product and $\Fb_L$ the
one-dimensional Fourier matrix that is a complex,
unitary, and symmetric matrix whose elements
are given by
\begin{equation}
(F_L)_{ij} = {\rm e}^{-2 \pi \iota (i-1)(j-1) /L},
\end{equation}
$N$ and $M$ are, respectively, the number of rows and columns of the image, $\fmatb^*$ is the complex conjugate
transpose of $\fmatb$, and $\Sigmab$
and $\Deltab$ are diagonal matrices containing the eigenvalues
of $\Hb$ and $\Lb$, respectively.  Note that if $\Lb = \Ib$, then
$\Deltab = \Ib$.  The eigenvalues can be obtained by
computing a two-dimensional fast Fourier transform (FFT) of the first
columns of $\Hb$ and $\Lb$, at a cost of $O(N^2 \log N)$, assuming
the blurred image contains $N \times N$ pixels. Recently,
however, many other efficient methods have been proposed that are suited to deal with a large variety of situations.

\subsection{Symmetric PSF}
If the spatially invariant PSF $h(x,y)$ is symmetric, but not necessarily separable,
i.e., $h(x,y) = h(-x,y) = h(x,-y)= h(-x,-y)$, it happens that also
the matrix $\Hb$ is symmetric. In this situation, it
is possible to show \citep{NgChTa99} that,
under reflexive boundary conditions,
the matrices $\Hb$ and $\Lb$ have the spectral
factorizations
\begin{equation}
\label{eq:Rfactor}
  \Hb = \cmatb^T \Sigmab \cmatb\,, \quad
  \Lb = \cmatb^T \Deltab \cmatb\,,
\end{equation}
where $\cmatb$ is the orthogonal two-dimensional
discrete cosine transform (DCT) matrix,
and $\Sigmab$
and $\Deltab$ are diagonal matrices containing the eigenvalues
of $\Hb$ and $\Lb$, respectively.
In this case, the eigenvalues of $\Hb$ are given by
\begin{equation}
  \sigma_i = \frac{\left[ \cmatb\Hb \eb_1 \right]_i}
                  {\left[ \cmatb\eb_1 \right]_i}\,,
\end{equation}
where
$\eb_1^T = \left[
           \begin{array}{cccc}
             1 & 0 & \cdots & 0
           \end{array}
         \right]$.
Note that $\Hb\eb_1$ is the first column of $\Hb$, which
can be constructed from the PSF, and that multiplication
by $\cmatb$ can be done in $O(N^2 \log N)$ operations
using fast DCT algorithms.  Computing the eigenvalues
$\delta_i$ of $\Lb$ is done in a similar manner.

To efficiently compute the regularization parameter, we first
replace $\Hb$ and $\Lb$ with their spectral factorizations
in Eq.~(\ref{eq:gcv1}), and simplify to obtain
\begin{equation} \label{eq:gcv}
  {\rm GCV}(\lambda)  =
                 n
                 \sum_{i=1}^{n}
                 \left(\frac{\hat{g}_i \delta_i^2}
                       {\sigma_i^2+\lambda^{2}\delta_i^2}
                 \right)^{2}
                  /
                  \left(
                         \sum_{i=1}^{n}
                         \frac{\delta_i^2}
                              {\sigma_i^2+\lambda^2 \delta_i^2}
                  \right)^2
\end{equation}
where, in the case of reflexive boundary conditions,
${\gbh}=\cmatb\gb$, and $n = N^2$ is the number of pixels in
the image.
We can now use standard minimization algorithms,
such as Newton's method,
to find the value of $\lambda$ that minimizes the scalar value function
(\ref{eq:gcv}).  In the computations reported in
this paper, we used Matlab's
{\tt fminbnd} function, which is based on Golden Section search
and parabolic interpolation.

We can also use the spectral factorization to efficiently
solve the least squares problem (\ref{eq:tikls}).  Because
$\cmat$ is an orthogonal matrix, (\ref{eq:tikls}) is
equivalent to
\begin{equation}
\label{eq:tikls2}
  \min \left\|
  \left( \begin{array}{c}
            \Sigmab \\ \lambda \Deltab
         \end{array} \right) \fbh -
  \left( \begin{array}{c}
            \gbh \\ \OO
         \end{array} \right)
      \right\|_2\,,
\end{equation}
where $\fbh = \cmatb\fb$ and $\gbh = \cmatb\gb$
(that is, $\fbh$ and $\gbh$ are, respectively,
DCTs of $\fb$ and $\gb$).
Because $\Sigmab$ and $\Deltab$ are diagonal matrices, this
least squares problem can be solved, using
a sequence of Givens rotations, with only $O(N^2)$
operations \citep{Hansen97}. In fact, strategic Givens
rotations permits
to change the structure of matrix
\begin{equation}
\left(
\begin{array}{c}
\Sigmab \\
\lambda \Deltab
\end{array}
\right)
\longrightarrow
\left(
\begin{array}{cccc}
* &   &   &  \\
  & * &   &  \\
  &   & \ddots &  \\
  &   &   & * \\
* &   &   &  \\
  & * &   &  \\
  &   & \ddots &  \\
  &   &   & *
\end{array}
\right)
\end{equation}
(``$*$'' indicates a non-zero element) to
\begin{equation}
\left(
\begin{array}{cccc}
* &   &   &  \\
  & * &   &  \\
  &   & \ddots &  \\
  &   &   & * \\
  &   &   &  \\
  &   &   &  \\
  &   &   &  \\
  &   &   &
\end{array}
\right)
\end{equation}
which is a form more amenable for an efficient solution.

To summarize, after $\lambda$ is computed by finding
the minimum of the ${\rm GCV}$ function,
the Tikhonov solution (\ref{eq:tikhonov})
can be computed as follows:
\begin{itemize}
\item[$\bullet$]
Construct the first column of $\Hb$ from the PSF
\item[$\bullet$]
Use a fast DCT algorithm to compute $\Sigmab$.
\item[$\bullet$]
Use a fast DCT algorithm to compute $\Deltab$.
\item[$\bullet$]
Solve the least squares problem (\ref{eq:tikls2}) to
compute $\fbh$.
\item[$\bullet$]
Use a fast inverse DCT algorithm to compute $\fb$ from $\fbh$.
\end{itemize}
The total cost of this approach is $O(N^2 \log N)$, and
storage requirements are only $O(N^2)$ (i.e., the storage
required for an $N \times N$ image). Specific implementations used for the
experiments are available in the Matlab package {\em RestoreTools}
\citep{LeNaPe02} \footnote{Available at
http://www.mathcs.emory.edu/$\sim$nagy/ RestoreTools}.

We remark that the efficient implementation described above assumes a symmetric $\Hb$
and reflexive boundary conditions.  Of course
a similar approach can be implemented for
periodic boundary conditions, using FFTs in place
of DCTs.  However, if we want to use reflexive boundary
conditions, and $\Hb$ is not symmetric, then other
methods should be considered.

\subsection{Separable PSF}

If the PSF is separable,
then $\Hb$ can be decomposed into a Kronecker product,
\begin{equation}
  \Hb = \Ab \otimes \Bb
      = \left(
          \begin{array}{cccc}
            a_{11} \Bb & a_{12} \Bb & \cdots & a_{1n} \Bb \\
            a_{21} \Bb & a_{22} \Bb & \cdots & a_{2n} \Bb \\
            \vdots   & \vdots   &        & \vdots   \\
            a_{m1} \Bb & a_{m2} \Bb & \cdots & a_{mn} \Bb
          \end{array}
        \right)\,.
\end{equation}
The special block structure of Kronecker products can be
exploited is several ways \citep{Jain89,KaNa98a}.  In particular,
assuming the images are $N \times N$ arrays of pixel values,
then the following properties hold:
\begin{enumerate}
\item
The first important property is that we need only store
the $N \times N$ matrices $\Ab$ and $\Bb$, and do not need to construct
the $N^2 \times N^2$ matrix $\Hb$ explicitly.
\item
The $N^2 \times N^2$ linear system
\begin{equation}
  (\Ab \otimes \Bb)\fb = \gb
\end{equation}
is equivalent to the $N \times N$ matrix equation
\begin{equation}
  \Ab \Fb \Bb^T = \Gb\,,
\end{equation}
where the $N^2 \times 1$ vectors $\fb$ and $\gb$ are obtained through
a lexicographical ordering of the $N \times N$ arrays $\Fb$ and $\Gb$.
\item
The singular value decomposition (SVD) is a tool used for analyzing
and solving ill-posed problems \citep{Hansen97}.  It is usually
too expensive for large scale problems, such as image deblurring.
However, for Kronecker product structures, the SVD can be used
efficiently.  To see this, suppose
\begin{equation}
  \Ab = \Ub_a \Sigmab_a \Vb_a^T \quad \mbox{and} \quad
  \Bb = \Ub_b \Sigmab_b \Vb_b^T
\end{equation}
are the SVDs of $\Ab$ and $\Bb$.  Then
\begin{equation}
  \Ab \otimes \Bb = (\Ub_a \otimes \Ub_b)(\Sigmab_a \otimes \Sigmab_b)
                    (\Vb_a^T \otimes \Vb_b^T)
\end{equation}
is the SVD of $\Hb = \Ab \otimes \Bb$.  In particular, the
SVD of a Kronecker product can be computed at a cost of only $O(N^3)$,
rather than $O(N^6)$ if working directly with $\Hb$.
\end{enumerate}
Using these properties, it looks like the SVD can be
used in place of the the spectral factorization of $\Hb$.
However, we run into difficulty if the regularization
operator, $\Lb$, is not the identity matrix.  To see this,
assume that the SVD of $\Hb$ is given by
\begin{equation}
  \Hb = \Ub_H \Sigmab \Vb_H^T\,,
\end{equation}
where $\Ub_H = \Ub_a \otimes \Ub_b$, $\Sigmab = \Sigmab_a \otimes \Sigmab_b$,
and $\Vb_H = \Vb_a \otimes \Vb_b$.  In addition, assume
$\Lb$ has the SVD
\begin{equation}
  \Lb = \Ub_L \Deltab \Vb_L^T\,,
\end{equation}
where the matrices $\Ub_H$, $\Vb_H$, $\Ub_L$, and $\Vb_L$
are orthogonal, and $\Sigmab$ and $\Deltab$ are diagonal.
If $\Lb = \Ib$, then $\Deltab = \Ib$, and
we can take $\Ub_L = \Vb_L = \Vb_H$.  Now Eqs.~(\ref{eq:gcv}) and (\ref{eq:tikls2})
can be used with
$\fbh = \Vb_H^T\fb$ and $\gbh = \Ub_H^T\gb$.
In this case, by the above properties
of Kronecker products, we see that the cost
of transforming (\ref{eq:gcv1}) into (\ref{eq:gcv})
and (\ref{eq:tikls}) into (\ref{eq:tikls2}) is $O(N^3)$.
This is slightly more expensive than the $O(N^2 \log N$)
cost when using fast transforms, but it is still a
reasonably efficient approach.  The storage requirements
remain $O(N^2)$.

A problem arises, though, when trying to use
a differentiation operator for $\Lb$ because, in general,
$\Vb_H \neq \Vb_L$.  If these two orthogonal matrices are
not equal, then (\ref{eq:tikls}) cannot be efficiently
transformed into (\ref{eq:tikls2}).  Moreover, (\ref{eq:gcv1})
cannot be efficiently transformed into (\ref{eq:gcv}), and
so evaluation of the ${\rm GCV}$ function, and hence computation
of its minimum, becomes much more expensive.  In this
difficult situation, a hybrid iterative/direct
method \citep{KiOL99} may be appropriate.

Finally, we remark that in general there
is no computational difference between
separable spatially variant PSFs
and separable spatially invariant blurs.  That is,
efficient direct methods can be used if $\Lb = \Ib$,
but not when using a differentiation operator for $\Lb$
\citep{KaNa98b}.  Moreover, in the difficult cases
when the direct factorization approach cannot be used,
iterative and hybrid methods can be implemented very
efficiently for both spatially invariant and spatially varying
blurs \citep{HaNa96,NaOL97,NaOL98}.

\subsection{Non-separable PSF}

In this section we have so far described efficient implementations
of direct methods for the following situations:
\begin{enumerate}
\item
Spatially invariant PSF, with periodic boundary conditions.  In this
case FFTs are used.
\item
Spatially invariant and symmetric PSF, with reflexive boundary conditions.
In this case fast cosine transforms are used.
\item
Separable PSF, invariant or variant.  In this case efficient use
of the Kronecker product structure is exploited.
\end{enumerate}
If the image deblurring problem does not fit into one of these
categories, then some other approach should be used.  One
possible option is to simplify,
or approximate, the problem with one that does fit into one of the
above categories. For example, a symmetric approximation,
such as $(\Hb + \Hb^T)/2$, can be
used with reflexive boundary conditions so that the fast
cosine transform approach can be used. Other, recently developed
approaches, use more sophisticated approximation techniques.

In particular the Kronecker approximation method \citep{KaNa98a,KaNa98b}
is a technique that deserves some attention. The main idea of this method is to
approximate the non-separable matrix $\Hb$ with a separable matrix.
To see how this can be done, suppose $\Pb$ is an $n \times n$ image of the PSF,
with $p_{ij}$ denoting the center (origin) or the PSF.  If the PSF is separable,
then we can write
\begin{equation}
  \Pb = {\bf a}{\bf b}^T \quad \mbox{and} \quad
  \Hb = \Ab \otimes \Bb\,,
\end{equation}
where:
\begin{itemize}
\item
Zero boundary conditions imply $\Ab$ and $\Bb$ are Toeplitz
matrices defined by ${\bf a}$ and ${\bf b}$, respectively.
That is,
\begin{equation}
  \Ab = \left( \begin{array}{ccccc}
                 a_i  & \cdots &  a_1   &        &        \\
               \vdots & \ddots &        & \ddots &        \\
                 a_n  &        & \ddots &        &   a_1  \\
                      & \ddots &        & \ddots & \vdots \\
                      &        &   a_n  & \cdots &   a_i
             \end{array}
       \right)\,
\end{equation}
and
\begin{equation}
  \Bb = \left( \begin{array}{ccccc}
                 b_j  & \cdots &  b_1   &        &        \\
               \vdots & \ddots &        & \ddots &        \\
                 b_n  &        & \ddots &        &   b_1  \\
                      & \ddots &        & \ddots & \vdots \\
                      &        &   b_n  & \cdots &   b_j
             \end{array}
       \right)\,.
\end{equation}
\item
Reflexive boundary conditions imply $\Ab$ and $\Bb$ are
Toeplitz-plus-Hankel
matrices defined by ${\bf a}$ and ${\bf b}$, respectively.
That is,
\begin{equation}
  \Ab = \left( \begin{array}{ccccc}
                 a_i  & \cdots &  a_1   &        &        \\
               \vdots & \ddots &        & \ddots &        \\
                 a_n  &        & \ddots &        &   a_1  \\
                      & \ddots &        & \ddots & \vdots \\
                      &        &   a_n  & \cdots &   a_i
             \end{array}
       \right) +
       \left( \begin{array}{ccccc}
                 a_{i+1} & \cdots  & a_n     &                  \\
                 \vdots  & \rddots &         &                  \\
                   a_n   &         &         &   a_1    \\
                         &         & \rddots & \vdots   \\
                         &     a_1 & \cdots  &   a_{i-1}
             \end{array}
       \right)\,,
\end{equation}
and
\begin{equation}
  \Bb = \left( \begin{array}{ccccc}
                 b_j  & \cdots &  b_1   &        &        \\
               \vdots & \ddots &        & \ddots &        \\
                 b_n  &        & \ddots &        &   b_1  \\
                      & \ddots &        & \ddots & \vdots \\
                      &        &   b_n  & \cdots &   b_j
             \end{array}
       \right) +
       \left( \begin{array}{ccccc}
                 b_{j+1} & \cdots  & b_n     &                  \\
                 \vdots  & \rddots &         &                  \\
                   b_n   &         &         &   b_1    \\
                         &         & \rddots & \vdots   \\
                         &     b_1 & \cdots  &   b_{j-1}
             \end{array}
       \right)\,.
\end{equation}
\end{itemize}
If the PSF is not separable, then we could compute a rank-one approximation
of $\Pb$, and construct $\Ab$ and $\Bb$ as described above.  That is,
\begin{equation}
   \Pb \approx {\bf a}{\bf b}^T \quad \Rightarrow \quad
   \Hb \approx \Ab \otimes \Bb\,.
\end{equation}
With a proper weighting applied to $P$, one obtains
an optimal approximation of the form
\begin{equation}
  \min||\Hb - \Ab \otimes \Bb||_F\,,
\end{equation}
where $||\cdot||_F$ is the Frobenius norm
and the minimization is done over all Kronecker products $\Ab \otimes \Bb$.
This approximation can be computed in $O(n^3)$ operations;
see \cite{KaNa98a,KaNa98b}
for further details.

\begin{acknowledgements}
Research of J. G. N. was partially supported by the National Science Foundation under Grant DMS 00-75239.
\end{acknowledgements}

\end{document}